\documentclass[a4paper,fleqn]{cas-sc}

\usepackage{amsfonts}
\usepackage{amsmath}
\usepackage{amssymb}
\usepackage{amsthm}
\usepackage{mathtools}
\usepackage{appendix}
\usepackage{booktabs}

\usepackage{bm}
\usepackage{bbm}
\usepackage[dvipsnames]{xcolor}
\usepackage[authoryear]{natbib}
\usepackage{lineno}
\usepackage{enumitem}
\usepackage[T1]{fontenc}
\usepackage{multirow}
\usepackage{soul}
\usepackage[font=small]{caption}
\usepackage{cancel}
\usepackage{algorithmicx}
\usepackage{algorithm}
\usepackage{algpseudocode}
\usepackage{float}
\usepackage{tikz}
\usetikzlibrary{intersections}
\usetikzlibrary{positioning}

\usepackage{graphicx,url}

\newtheorem{theorem}{Theorem}

\newtheorem{corollary}{Corollary}

\usepackage[utf8]{inputenc}  

\graphicspath{ {../images/} }

\newcommand{\blind}{1}

\newcommand{\norm}[1]{\lVert#1\rVert}

\newcommand{\bt}{\textbf{t}}
\newcommand{\bv}{\textbf{v}}
\newcommand{\bbf}{\textbf{f}}
\newcommand{\rd}{\mathrm{d}}
\newcommand{\rX}{\mathrm{X}}

\newcommand{\bgamma}{\bm{\gamma}}

\newcommand{\btau}{\bm{\tau}}

\newcommand{\bomega}{\bm{\omega}}
\newcommand{\T}{\intercal}
\newcommand{\one}{\mathbbm{1}}

\begin{document}
\let\WriteBookmarks\relax

\shorttitle{Closed-form Covariance Functions in Space and Time}
\shortauthors{L. Astfalck}

\if1\blind
  \title[mode=title]{Closed-form Covariance Functions in Space and Time from B-spline Spectral Densities}
  \author[1]{Lachlan Astfalck}[orcid=0000-0002-8369-9384]
  \affiliation[1]{organization={School of Mathematics \& Statistics, University of New South Wales},
                  city={Kensington},
                  country={Australia}}
\fi

\if0\blind
  \title[mode=title]{Closed-form Covariance Functions in Space and Time from B-spline Spectral Densities}
\fi

\begin{abstract}
    The autocovariance function (ACF) is central to the analysis of time series, spatial, and spatio-temporal data. Classical parametric models are often too restrictive for multivariate and space-time data, which motivates non-parametric descriptions of the ACF. Bochner's Theorem guarantees that any positive spectral measure yields a valid ACF via the inverse Fourier transform; however, existing non-parametric approaches in the spectral domain rarely return closed-form expressions for the ACF itself. We develop a closed-form class of non-parametric ACFs by deriving the inverse Fourier transform of B-spline spectral bases with arbitrary degree and knot placement. This yields a family of closed-form stationary covariance models with three properties: (i) it is uniformly dense for the ACF, among the ACFs of weakly stationary processes that possess a spectral density; (ii) it accommodates univariate, multivariate, and multidimensional processes; and (iii) it is non-separable in general, with separability a special case. Jackson-type approximation bounds establish convergence rates, and simulation studies show accurate recovery of known univariate and bivariate ACFs. Applied to output from a high-resolution ocean simulation, the fit space-time ACF is non-separable and demonstrates accurate process recovery.
\end{abstract}


\maketitle

\section{Introduction} \label{sec:intro}

The autocovariance function (ACF) captures the second-order structure of a weakly stationary process and underpins much of modern time-series, spatial, and spatio-temporal analysis. Parametric models are widely used, with rich catalogues for univariate processes \citep{brockwell2002introduction,williams2006gaussian} and more nuanced constructions for multivariate and multidimensional settings \citep{stein2005space,cressie2011statistics}. In practice, however, model choice is often ad hoc, and the assumptions underlying parametric forms may be overly rigid for complex applications. Because the ACF must be positive semi-definite, constructing valid non-parametric ACFs in closed form remains a fundamental challenge. Recent work has addressed this by approximating the ACF via basis expansions of the power spectral density \citep[PSD;][]{wilson2013gaussian,parra2017spectral,tobar2019band,simpson2021minecraft}. Whilst these methods offer flexibility, they often impose restrictive structural assumptions on the basis functions. Approximation guarantees exist, but typically rely on bases that are suboptimal for functional representation, exhibiting slow convergence and poor approximation efficiency. In this paper, we propose a general class of closed-form, non-parametric ACF kernels that are (1) provably uniformly dense, among the ACFs of weakly stationary processes that possess a spectral density, and (2) capable of efficient approximation, with Jackson-type rates of convergence to the target PSD. We achieve this by constructing spline-based expansions of arbitrary smoothness for the PSD and analytically deriving their inverse Fourier transforms. In this setting, non-separability is handled naturally: it emerges from the model structure rather than being imposed. We prove that this class is dense among weakly-stationary multivariate and multidimensional ACFs that admit a spectral density, providing a principled and scalable foundation for non-parametric modelling of the ACF.

Let ${\rX(t)}$ be a stationary random process with ACF $\bgamma(\btau) \in \mathbb{C}^{M \times M}$ for $\btau \in \mathbb{R}^D$, where $M, D \in \mathbb{N}^+$. We refer to the process as \emph{univariate} when $M = D = 1$, \emph{multivariate} when $M > 1$, and \emph{multidimensional} when $D > 1$. Throughout, we use boldface $\bgamma$ and $\btau$ for multivariate and multidimensional settings, and $\gamma$ and $\tau$ otherwise. We allow $\bgamma(\btau)$ to be complex-valued in general. The literature on parametric modelling of univariate ACFs is extensive, and we defer to one of the many book-length treatments for a detailed review \citep[e.g.][]{williams2006gaussian}. Despite the depth of research, there is no parametric family that captures all valid forms of $\gamma(\tau)$. In practice, complex ACFs are often constructed from sums and products of simpler components; but, this remains a modelling choice rather than a principled derivation. When $M > 1$, the function $\bgamma(\tau)$ must be blockwise positive semi-definite, satisfying $\sum_{i,j} \bv_i^\mathrm{H} \bgamma(t_i {-} t_j) \bv_j \geq 0,$ for all $\{\bv_i\} \subset \mathbb{C}^M$, where the superscript $\mathrm{H}$ denotes the Hermitian transpose. This constraint can be prohibitive even for relatively simple forms. For instance, generalising the Matérn ACF to the multivariate setting requires considered specification of the parameters controlling the cross-covariance structure \citep{gneiting2010matern,apanasovich2012valid}. When $D > 1$, as for random fields or for spatio-temporal processes, the most common approach to defining $\gamma(\btau)$ is to assume separability, i.e., $\gamma(\btau) = \gamma_1(\tau_1) \cdots \gamma_D(\tau_D)$, where each marginal $\gamma_i(\tau_i)$ is an ACF. While computationally convenient \citep{genton2007separable}, this assumption is often too restrictive in practice \citep{stein2005space,fuentes2008class}. Non-separable covariance classes do exist \citep[e.g.][]{cressie1999classes, gneiting2002nonseparable}; however, as in the multivariate case, there is no general methodology for constructing analytical representations of non-separable ACFs.

Such functional flexibility can be achieved for multivariate and space-time dependence by utilising non-parametric methods directly in the spectral domain. \citet{im2007semiparametric} model an isotropic spectral density as a cubic B-spline expansion on a uniform partition below a cutoff frequency, joined continuously to a truncated algebraic tail. \citet{kleiber2017coherence} develops coherence, phase and gain for multidimensional stationary processes, estimating them from smoothed periodogram matrices. Closest to our own work, \citet{qadir2021semiparametric} expand the coherence function between approximately Mat\'{e}rn marginals in cubic B-splines and give sufficient conditions for the resulting matrix-valued covariance to be valid. The representations induced by their methodology have no known closed form and are evaluated by a finite sum over a discretised frequency grid. Finally, \citet{yarger2026multivariate} and \citet{yarger2026connecting} obtain parametric families of asymmetric and non-separable cross-covariances from a stochastic integral representation. In general, these methodologies do not allow for a closed-form ACF for a non-parametric spectral density of arbitrary differentiability.

A more recent line of research proposes closed-form non-parametric estimates of the ACF via inverse Fourier transforms. Bochner's theorem states that $\bgamma(\btau)$ is a valid ACF of a mean-zero, weakly stationary process if and only if it is the inverse Fourier transform of a finite, positive measure. When the spectral measure admits a density this yields the identity
\begin{equation} \label{eqn:bochner}
  \bgamma(\btau) = \int_{\mathbb{R}^D} \bbf(\bomega) e^{2\pi \iota \btau^\T \bomega} \, \rd \bomega \quad \Longleftrightarrow \quad \bbf(\bomega) = \int_{\mathbb{R}^D} \bgamma(\btau) e^{-2\pi \iota \btau^\T \bomega} \, \rd \btau,
\end{equation}
where $\iota$ denotes the imaginary unit, and $\bbf(\bomega)$ is the PSD. For real-valued processes, $\bbf(-\bomega) = \bbf(\bomega)^\T$, so that in the univariate case the PSD is symmetric, $f(-\omega) = f(\omega)$. \citet{wilson2013gaussian} exploit \eqref{eqn:bochner} by proposing a flexible basis expansion of the univariate PSD of the form
\[
f(\omega) = \sum_i \frac{\beta_i}{2}\left[\phi(\omega; \mu_i, \sigma_i^2) + \phi(-\omega; \mu_i, \sigma_i^2)\right],
\]
where $\beta_i \geq 0$ are weights and $\phi(\cdot; \mu_i, \sigma_i^2)$ denotes the Gaussian density with mean $\mu_i$ and variance $\sigma_i^2$. Whilst this construction ensures analyticity of the inverse Fourier transform, and hence the ACF, the infinite support and continuity of the Gaussian density may be undesirable. \citet{tobar2019band} address this by constructing a band-limited representation of the form
\[
f(\omega) = \sum_i \frac{\beta_i}{2}\left[\Pi(\omega; c_i, w_i) + \Pi(-\omega; c_i, w_i)\right],
\]
where $\beta_i \geq 0$ are weights and $\Pi(\omega; c_i, w_i)$ is a rectangular (boxcar) function centered at $c_i$ with width $w_i$. This construction corresponds to a spline basis of degree zero, and hence yields a piecewise-constant, but discontinuous, representation of the PSD. These ideas have been extended to the multivariate case. \citet{parra2017spectral} generalised the Gaussian basis of \citet{wilson2013gaussian}, but the resulting basis is not dense in the space of valid matrix-valued ACFs \citep{simpson2021minecraft}. Similarly, \citet{simpson2021minecraft} extended the band-limited construction of \citet{tobar2019band} to multivariate settings, but the resulting PSDs remain discontinuous and non-differentiable. A rich literature exists on non-parametric estimation of the PSD \citep[e.g.][]{choudhuri2004bayesian,rosen2007automatic,rosen2012adaptspec,edwards2019bayesian,astfalck2024debiasing,astfalck2025bias}, although rarely do these approaches admit a closed-form ACF.

The current literature, in effect, presents two extremes for non-parametric representations of the ACF via a spectral basis expansion: those that are infinitely differentiable \citep[see][]{wilson2013gaussian}, and those that are non-differentiable \citep[see][]{tobar2019band}. Neither provides a principled mechanism for controlling smoothness, which limits their approximation efficiency and flexibility. We address these limitations by studying B-splines, a classical tool in approximation theory that offers tunable smoothness, compact support and interpretability. This leads to a generalised and universal construction: a spline basis of arbitrary degree for the PSD, parametrised by the knots \(\mathcal{K} = \{\kappa_0 , \kappa_1 , \dots , \kappa_{m+k}\}\). In the univariate setting, our main result is the derivation of closed-form expressions for the inverse Fourier transform under general (non-uniform) knot placements. Theoretical properties are established through a Jackson-type inequality that bounds the $L^q$-norm between the true and modelled ACFs, and characterises the convergence rate as a function of the smoothness of the PSD, the spline degree \(k\) and the maximal knot spacing. These results extend naturally to the multivariate setting ($ M > 1 $). We prove that the space of matrix-valued ACFs generated by our method is dense in the $L^1$ metric on spectral densities, and hence uniformly for the ACF, for every degree $k$ (Theorem~\ref{the:multivariate}); in contrast, the Gaussian basis of \citet{parra2017spectral} cannot represent the full range of cross-spectra between components \citep{simpson2021minecraft}. The piecewise-constant basis is also dense \citep{simpson2021minecraft}; however, its error bound has no decay in $\tau$ (Corollary~\ref{cor:acf_tail_bound}), whilst for $k \geq 1$ we achieve a tail-error bound of $\mathcal{O}(|\tau|^{-k})$. Further, we extend our construction to the multidimensional setting ($ D > 1 $), where we obtain a flexible framework for modelling non-separable dependencies. We demonstrate the utility of our method on simulation studies of univariate and multivariate processes. Finally, we apply our methodology to model the ACF of a complex multidimensional space-time field of the ocean. Our model captures non-separable structure without imposing parametric assumptions or separability constraints.

The remainder of this paper is structured as follows. Section~\ref{sec:univariate} presents results for univariate random processes. Sections~\ref{sec:multivariate} and~\ref{sec:multidimensional} extend these results to the multivariate (\(M > 1\)) and multidimensional (\(D > 1\)) settings, respectively. Simulation studies of univariate and multivariate random processes are shown in Section~\ref{sec:simulation}. In Section~\ref{sec:applications}, we demonstrate the method on a numerical model output of an oceanographic flow,  and discuss some relevant computational considerations. Section~\ref{sec:discussion} concludes. Proofs of all theorems are deferred to the Appendix. We now provide the assumptions that underpin the theoretical results in this paper. 

\subsection{Assumptions} \label{sec:assumptions}

We consider a mean-zero, weakly stationary, mean-square continuous random process $\{\rX(t)\}_{t \in \mathbb{R}^D}$ taking values in $\mathbb{C}^M$. Allowing $\rX(t)$ to be complex-valued simplifies the mathematical developments whilst providing a more general solution. The autocovariance function $\bgamma(\btau)$ is assumed to exist and be given by the inverse Fourier transform of a matrix-valued spectral density $\bbf : \mathbb{R}^D \to \mathbb{C}^{M \times M}$, where  $\bbf(\bomega)\succeq 0$ is positive semi-definite and Hermitian for all $\bomega$. We assume that $\bbf \in L^1(\mathbb{R}^D) \cap \mathcal{W}^{r,p}(\mathbb{R}^D)$, where $\mathcal{W}^{r,p}(\mathbb{R}^D)$ is the Sobolev space of functions with weak derivatives up to order \(r \geq 1\) in \(L^p(\mathbb{R}^D)\) for some \(1 \leq p \leq 2 \). This induces a class of ACFs that vanishes at infinity and precludes the presence of Dirac spikes in $\bbf$, corresponding to non-decaying harmonic signals. As $\bbf \in L^p(\mathbb{R}^D)$, $\bbf$ is compactly supported to arbitrary precision; that is, for all $\varepsilon_a > 0$, there exists some $0 \leq a < \infty$ such that the truncated function $\bbf_{[-a,a]^D}(\bomega) := \bbf(\bomega) \mathbf{1}_{[-a,a]^D}(\bomega)$ satisfies $\|\bbf - \bbf_{[-a,a]^D}\|_{L^p(\mathbb{R}^D)} < \varepsilon_a$. For multivariate $\bbf$, all derivatives and norms are interpreted entry-wise.

\section{Univariate Random Processes} \label{sec:univariate}

Let \(\mathcal{S}_{k,\mathcal{K}}\) denote the space of piecewise polynomial functions of degree at most \(k\), defined on a strictly increasing knot sequence \(\mathcal{K} = \{\kappa_0, \kappa_1, \dots, \kappa_{m+k}\}\). Each function \(S \in \mathcal{S}_{k,\mathcal{K}}\) is a polynomial of degree at most \(k\) on each interval \([\kappa_i, \kappa_{i+1}]\), and has \(k - 1\) continuous derivatives globally. This space is spanned by a B-spline basis \citep{de1978practical}, denoted \(\{B_{i,k}(\omega)\}_{i=0}^{m-1}\), where each basis function is compactly supported on a subinterval of \(\mathcal{K}\). The B-splines span the space of such piecewise polynomials on \([\kappa_k, \kappa_m]\). The most common definition of B-splines is given via the Cox--de Boor recursion,
\begin{equation*}
  B_{i,0}(\omega) = 
  \begin{cases}
    1 & \quad  \kappa_{i} \leq \omega < \kappa_{i+1}, \\
    0 & \quad \text{otherwise},
  \end{cases}
\end{equation*}
and for \(k \geq 1\),
\begin{equation} \label{eqn:spline_definition}
  B_{i,k}(\omega) = \frac{\omega - \kappa_i}{\kappa_{i+k} - \kappa_i} B_{i,k-1}(\omega) + \frac{\kappa_{i+k+1} - \omega}{\kappa_{i+k+1} - \kappa_{i+1}} B_{i+1,k-1}(\omega).
\end{equation}
As seen in \eqref{eqn:spline_definition}, each B-spline basis function is uniquely determined by its degree \(k\) and the knot sequence \(\mathcal{K}\), and has \(k - 1\) continuous derivatives.

The foundational idea of this manuscript is that, to an arbitrary order of differentiability and resolution in $\omega$, we can obtain a non-parametric description of the PSD from a B-spline representation
\begin{equation} \label{eqn:psd_basis}
  \hat{f}_k(\omega) = \sum_{i=0}^{m-1} c_i B_{i, k}(\omega)
\end{equation}
with coefficients $c_i$ ensuring $\hat{f}_k(\omega) \geq 0$ for all $\omega$; Section~\ref{sec:nonnegativity} documents how this constraint is imposed in practice. From \eqref{eqn:psd_basis}, a non-parametric description of the ACF, $\hat{\gamma}_k(\tau)$, is readily available in closed-form via the inverse Fourier transform of $\hat{f}_k(\omega)$. This is formalised in the following Theorem.

\begin{theorem} \label{the:acf_closed_form}
  Let $\{B_{i,k}(\omega)\}_{i=0}^{m-1}$ denote a B-spline basis of degree $k$ over an ordered knot sequence $\mathcal{K} = \{\kappa_0,  \dots, \kappa_{m+k}\}$, and let $\hat{f}_k(\omega) = \sum_{i=0}^{m-1} c_i B_{i,k}(\omega)$ be a linear combination of B-splines such that $\hat{f}_k(\omega) \geq 0$. Define $\rho_{i,k}(\tau)$ as the inverse Fourier transform of $B_{i,k}(\omega)$. Then, via \eqref{eqn:bochner}, the associated autocovariance function
  \begin{equation} \label{eqn:acf_hat}
  \hat{\gamma}_k(\tau) = \int_{-\infty}^\infty \hat{f}_k(\omega) e^{2\pi \iota \omega \tau} \, \rd \omega = \sum_{i=0}^{m-1} c_i \rho_{i,k}(\tau)
  \end{equation}
  is positive semi-definite and admits a closed-form expression. In particular, the ACF basis function $\rho_{i,k}(\tau)$ is given by
  \begin{equation} \label{eqn:non_uniform_acf_basis}
  \rho_{i,k}(\tau) = \frac{(-1)^k k!}{(2\pi \iota \tau)^{k+1}} \sum_{j=i}^{i+k} \alpha_j e^{2\pi \iota \kappa_j \tau} \left(
  e^{2\pi \iota \kappa_{\mathrm{w},j} \tau} \sum_{l=0}^k \frac{(-2\pi \iota \kappa_{\mathrm{w},j} \tau)^l}{l!} - 1
  \right),
  \end{equation}
  where $\kappa_{\mathrm{w},j} := \kappa_{i+k+1} - \kappa_j$, and the $\alpha_j$ are the coefficients from the shifted truncated power basis representation of $B_{i,k}(\omega)$, defined below in Section~\ref{sec:non_uniform}. Equation \eqref{eqn:non_uniform_acf_basis} holds for $\tau \neq 0$, and $\rho_{i,k}(0) = \int B_{i,k}(\omega) \, \rd \omega = (\kappa_{i+k+1} - \kappa_i)/(k+1)$.
  \end{theorem}
  
The proof of this Theorem is outlined in Section~\ref{sec:non_uniform} with more tedious details in the Appendix. When the knots are uniformly spaced, Theorem~\ref{the:acf_closed_form} simplifies to a well-known result, which we highlight in the next section through a simplified example that clarifies the core concepts.

\subsection{Example with uniform knot placement}

Assume the knots are spaced uniformly so that $\kappa_i - \kappa_{i-1} = h$ for all $i$. The ACF basis corresponding to a zeroth-degree spline, $\rho_{i, 0}(\tau)$ is
\begin{equation} \label{eqn:rho0_even}
  \rho_{i,0}(\tau) = h e^{2 \pi \iota (\kappa_i + h/2) \tau} \mathrm{sinc}(h \tau)
\end{equation}
where $\mathrm{sinc}(x) = \sin(\pi x)/(\pi x)$ denotes the normalised sinc function, with $\mathrm{sinc}(0) = 1$. Here, we have assumed the general case of complex valued $\rho_{i,0}(\tau)$;  in practice the data are usually real-valued, in which case the ACF is real and even and the spectral density is symmetric. The basis can be symmetrised via $\frac{1}{2}[B_{i,k}(\omega) + B_{i,k}(-\omega)]$; in this example, yielding $\rho_{i,0}(\tau) = h \cos\{2 \pi (\kappa_i + h/2) \tau\} \mathrm{sinc}(h \tau)$. In the case of equal knot spacing, the recursion in \eqref{eqn:spline_definition} is given by the convolution $B_{i, k}(\omega) = h^{-1} (B_{i, 0} \ast B_{i, k-1})(\omega + \kappa_i)$ and so via the convolution theorem, we obtain the ACF basis that corresponds to a general $k$th degree spline as
\begin{equation} \label{eqn:rho_equal_space}
  \rho_{i, k}(\tau) = h \left\{\mathrm{sinc}(h \tau)\right\}^{k+1} e^{2 \pi \iota \{\kappa_i + (k+1)h/2\} \tau}.
\end{equation}
This is a special case of \eqref{eqn:non_uniform_acf_basis} in Theorem~\ref{the:acf_closed_form}, for when $\kappa_i - \kappa_{i-1} = h$ for all $i$.

\begin{figure}[pos=H]
  \centering
  \includegraphics[width=0.7\textwidth]{"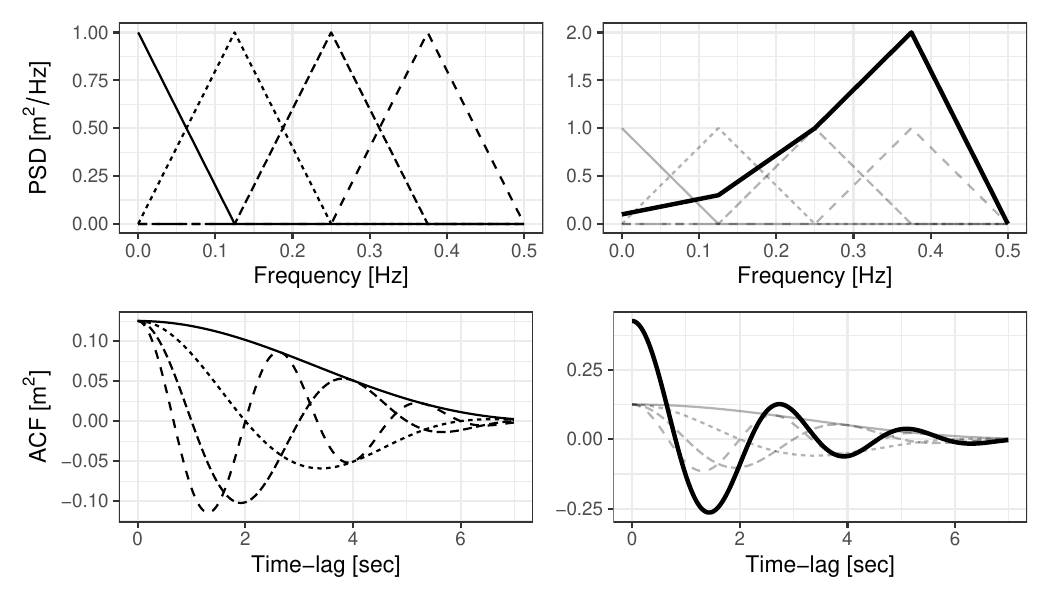"}
  \caption{
    Illustration of B-spline basis functions and their corresponding ACF bases.
    Top left, linear B-spline bases \( B_{i,1}(\omega) \) with uniformly spaced knots \(\mathcal{K} =  \{-0.125, 0, 0.125, 0.25, 0.375, 0.5\}\).
    Bottom left, corresponding inverse Fourier transforms \( \rho_{i,1}(\tau) \).
    Top right, a weighted basis combination \( \hat{f}_1(\omega) = \sum_{i=0}^3 c_i B_{i,1}(\omega) \) using coefficients \( \{c_0, \dots, c_3\} = \{0.1, 0.3, 1, 2\} \), the individual spline bases from the top left panel are shown in grey.
    Bottom right, the resulting autocovariance function \( \hat{\gamma}_1(\tau) \), the individual ACF bases from the bottom left panel are shown in grey.
    All quantities assume $\rX(t)$ measured in metres and $t$ in seconds.
  }
  \label{fig:uniform}
\end{figure}

We show a simple depiction of the non-parametric ACF representation in Figure~\ref{fig:uniform}. Define the bases $B_{i,1}(\omega)$ with knot sequence $\mathcal{K} =  \{-0.125, 0, 0.125, 0.25, 0.375, 0.5\}$, shown in the top left panel. The real component of the corresponding $\rho_{i,1}(\tau)$, via \eqref{eqn:rho_equal_space}, are shown in the bottom left panel. Say some $\hat{f}_1(\omega)$, that approximates a true $f(\omega)$, is constructed from the coefficient values $\{c_0, \dots, c_3\} = \{0.1, 0.3, 1, 2\}$, shown by the bold line in the top right panel. The corresponding $\hat{\gamma}_1(\tau) = \sum_{i=0}^3 c_i \rho_{i,1}(\tau)$, which is an estimate of the true $\gamma(\tau)$, is shown in the bottom right panel. In order to attribute units to the plot, we assume that $\rX(t)$ is measured in metres and $t$ in seconds.

\subsection{Derivation of closed-form representation for general knots} \label{sec:non_uniform}

The previous example gives a convenient form for when the knots are uniformly spaced. A key advantage of our methodology is that the knot placement need not be uniform, and may be adaptively designed to best fit the structure of the PSD. We now motivate a proof to Theorem~\ref{the:acf_closed_form}; more tedious inverse Fourier calculations are deferred to the Appendix.
When the knot placement is non-uniform, the convolution representation to \eqref{eqn:spline_definition} does not hold and so the representation of $\rho_{i, k}(\tau)$ in \eqref{eqn:rho_equal_space} is not valid in the general case. To obtain a general solution, we seek an alternative form to \eqref{eqn:spline_definition}. Define the shifted truncated power function as
\begin{equation*}
  (\omega - \kappa_j)^k_+ = \begin{cases}
    (\omega - \kappa_j)^k & \quad \omega \geq \kappa_j \\
    0 & \quad \mathrm{otherwise}.
  \end{cases}
\end{equation*}
As shown in \cite{de1978practical}, the representation in \eqref{eqn:spline_definition} may be alternatively given as 
\begin{equation} \label{eqn:tpb}
  B_{i,k}(\omega) = \sum_{j=i}^{i+k+1} \alpha_j (\omega - \kappa_j)^k_+,
\end{equation}
where the coefficients $\alpha_j$ depend on the knot placements and are solved so that \eqref{eqn:spline_definition} and \eqref{eqn:tpb} are equivalent. We provide closed-form solutions for linear and quadratic splines in the Appendix, together with a divided-difference expression valid for any degree \citep[see][]{de1978practical}. The shifted truncated power function does not have a finite inverse Fourier transform as it is unbounded as $\omega \rightarrow \infty$. However, noting that the role of the $(i+k+1)$th basis in \eqref{eqn:tpb} is to enforce $B_{i,k}(\omega) = 0$ for $\omega > \kappa_{i+k+1}$, we may equivalently define \eqref{eqn:tpb} as
\begin{equation} \label{eqn:double_truncation}
  B_{i,k}(\omega) = \sum_{j=i}^{i+k} \alpha_j (\omega - \kappa_j)^k_+ \one\{\omega < \kappa_{i+k+1}\}.
\end{equation}
Thus, the inverse Fourier transform of the $j$th truncated power basis in \eqref{eqn:double_truncation} is
\begin{align}
  \int_{-\infty}^\infty \one\{\omega < \kappa_{i+k+1}\} (\omega - \kappa_j)^k_+& e^{2 \pi \iota \omega \tau} \; \rd \omega = \int_{\kappa_j}^{\kappa_{i+k+1}} (\omega - \kappa_j)^k e^{2 \pi \iota \omega \tau} \; \rd \omega \nonumber \\
  &= \frac{(-1)^k \cdot k!e^{2\pi \iota \kappa_j \tau}}{(2\pi \iota \tau)^{k+1}} \left[
    e^{2\pi \iota \kappa_{\mathrm{w},j} \tau} \cdot \sum_{l=0}^k \frac{(-2\pi \iota \kappa_{\mathrm{w},j} \tau)^l}{l!} - 1 \label{eqn:j-tpb}
    \right],
\end{align}
where, here, $\kappa_{\mathrm{w},j} = \kappa_{i+k+1} - \kappa_{j}$ is the domain of the integration, and a full derivation is in the Appendix. The summation in \eqref{eqn:j-tpb} is over the polynomial order $k$ and is easy to compute in practice.

\subsection{Jackson-type Convergence}

Jackson's inequality provides a bound on the approximation error between a spline estimator and the true underlying function, in this case, the PSD. Using the Hausdorff-Young inequality, we can translate these statements into convergence bounds on the true ACF $\gamma$ and the spline approximated version $\hat{\gamma}_k$. Here, we state a Jackson-type inequality, and demonstrate implications via a series of corollaries, see the Appendix for proof. Multivariate and multidimensional extensions to all statements exist.

\begin{theorem} \label{the:jackson}
  Let $f \in L^1(\mathbb{R}) \cap \mathcal{W}^{r, p}(\mathbb{R})$, where $\mathcal{W}^{r,p}(\mathbb{R})$ is the Sobolev space of $r$-times weakly differentiable functions in $L^p(\mathbb{R})$, for some $1 \leq p \leq 2$ and some integer $1 \leq r \leq k+1$, and $f$ is compactly supported to arbitrary precision so that for every $\varepsilon_a > 0$, there exists some $a > 0$ such that $\|f\|_{L^p(\mathbb{R} \setminus [-a,a])} < \varepsilon_a$.  Let $\{\kappa_0 < \dots < \kappa_{m+k}\}$ be a knot sequence whose span contains $[-a, a]$, with $k$ knots beyond each end, so that $\kappa_k = -a$ and $\kappa_m = a$, with maximum and minimum knot distances $h_{\max} = \max_i\{\kappa_{i} - \kappa_{i-1}\}$ and $h_{\min} = \min_i\{\kappa_{i} - \kappa_{i-1}\}$, whose ratio $h_{\max}/h_{\min}$ remains bounded as the knots are refined. Then there exists a spline approximation $\hat{f}_k \in \mathcal{S}_{k,\mathcal{K}}$ such that the autocovariance function $\gamma(\tau)$ and its spline-based approximation $\hat{\gamma}_k(\tau)$, from Theorem~\ref{the:acf_closed_form}, satisfy
  \begin{equation*}
    \norm{\gamma - \hat{\gamma}_k}_{L^q(\mathbb{R})} \leq C_{k,p} \big( h^{r}_{\max} \norm{f^{(r)}}_{L^p([-a,a])} + \varepsilon_a \big), \quad \text{for} \quad \frac{1}{p} + \frac{1}{q} = 1,
  \end{equation*}
  where $f^{(r)}$ is the $r$th derivative of $f$, $\varepsilon_a$ can be made arbitrarily small, and $C_{k,p}$ is a constant depending only on $k$, $p$ and the bound on $h_{\max}/h_{\min}$. Density of the valid class, with $\hat{f}_k \geq 0$, follows from Theorem~\ref{the:multivariate} with $M = 1$.
\end{theorem}

This result shows that the approximation error of the ACF decays at a rate determined by the smoothness $r$ of the PSD, the maximum knot spacing $h_{\max}$, and the regularity of $f$ via its $r$th derivative. While the smoothness $r$ and the size of $\|f^{(r)}\|_{L^p}$ are intrinsic to the underlying process and not directly controllable, the overall approximation error can be reduced to an arbitrary degree by refining the knot spacing. The degree $k$ is chosen, and raising it improves the rate only while $k + 1 \leq r$. By choosing $p=1$ and $q = \infty$, we can establish uniform convergence of $\hat{\gamma}_k(\tau)$ to $\gamma(\tau)$.

\begin{corollary}
  Theorem~\ref{the:jackson} implies uniform convergence of the ACF. Specifically, if $f \in \mathcal{W}^{r,1}(\mathbb{R})$, setting $p = 1$ in Theorem~\ref{the:jackson} (or simply via the triangle inequality), 
  \begin{equation*}
    \sup_{\tau \in \mathbb{R}} |\gamma(\tau) - \hat{\gamma}_k(\tau)| \leq \norm{f - \hat{f}_k}_{L^1(\mathbb{R})} \leq C_{k,1} \big( h^{r}_{\max} \norm{f^{(r)}}_{L^1([-a,a])} + \varepsilon_a \big),
  \end{equation*}
  providing a uniform bound on the difference between the true and approximated ACFs.
\end{corollary}

Uniform convergence of the ACF allows for distributional convergence of the underlying stochastic processes. So long as the finite-dimensional distributions of $\rX$ are determined by $\gamma$, as for a Gaussian process (GP), uniform convergence of the ACF is sufficient to guarantee convergence of the finite-dimensional distributions.

\begin{corollary}
  The uniform convergence of \( \hat{\gamma}_k \rightarrow \gamma \), together with the assumption that the finite-dimensional distributions of the processes \( X \) and \( \hat{X}_k \) are determined by their autocovariance functions, implies convergence of the finite-dimensional distributions of the approximating processes \( \hat{X}_k \) to \( X \). In particular, this condition is satisfied if \( X \) and \( \hat{X}_k \) are Gaussian processes.
\end{corollary}

Finally, we provide a bound on the tail error of the ACF approximation, establishing a decay rate of $|\gamma(\tau) - \hat{\gamma}_k(\tau)| = \mathcal{O}( |\tau|^{-k})$. This result quantifies how smoothness of the spectral density translates to long-lag accuracy in the autocovariance domain.

\begin{corollary}
\label{cor:acf_tail_bound}
Let $f \in W^{r,1}(\mathbb{R})$, and let $\hat{f}_k \in \mathcal{S}_{k,\mathcal{K}}$ be a spline approximation of degree $k$, with the assumptions of Theorem~\ref{the:jackson} and the additional condition that $\|f^{(j)}\|_{L^1(\mathbb{R} \setminus [-a,a])} \leq \varepsilon_a$ for all $j < r$. For any $j < r$, via derivative properties of the Fourier transform, the autocovariance function $\gamma(\tau)$ and its spline-based approximation $\hat{\gamma}_k(\tau)$ satisfy
\[
|\gamma(\tau) - \hat{\gamma}_k(\tau)| \leq \frac{1}{(2\pi |\tau|)^j} \norm{f^{(j)} - \hat{f}^{(j)}_k}_{L^1(\mathbb{R})}.
\]
Using the extension of Jackson's inequality to derivatives \citep[see][Chapter~6]{schumaker2007spline},
\[
\norm{f^{(j)} - \hat{f}^{(j)}_k}_{L^1([-a,a])} \leq C h_{\max}^{r-j} \| f^{(r)} \|_{L^1([-a,a])},
\]
and we obtain the global tail bound
\begin{equation*}
  |\gamma(\tau) - \hat{\gamma}_k(\tau)| 
  \leq \frac{C}{(2\pi |\tau|)^j} h_{\max}^{r-j} \| f^{(r)} \|_{L^1([-a,a])} + \frac{\varepsilon_a}{(2\pi |\tau|)^j}.
\end{equation*}
Choosing \( r = k+1 \) and \( j = k \) we achieve the bounded tail decay error of $|\gamma(\tau) - \hat{\gamma}_k(\tau)| = \mathcal{O}( |\tau|^{-k})$.
\end{corollary}

For $k = 0$ the corollary admits only $j = 0$ and gives no decay in $\tau$, corresponding to \citet{tobar2019band}; each increase in the degree, for a sufficiently smooth spectral density, adds one power of $|\tau|^{-1}$ to the tail bound. The effect of this is seen in Figure~\ref{fig:1d_acfs} in the simulation studies, below.

\subsection{Implications of a compactly supported spectral density} \label{sec:continuous_domain}

Every member of $\mathcal{S}_{k,\mathcal{K}}$ vanishes outside the knot span $[\kappa_0, \kappa_{m+k}]$, so every PSD in the class is compactly supported. Consequently $\hat{\gamma}_k$ is real-analytic \citep[Theorem~7.3.1]{hormander1990analysis}, describing a process that is mean-square infinitely differentiable. In contrast, a target ACF that is rough at the origin, such as the exponential or a Mat\'ern ACF of low smoothness, is therefore approximated to the accuracy of Theorem~\ref{the:jackson} but may never be exactly modelled. Compact support also has a numerical consequence in an in-fill regime. The covariance matrix of $\hat{\gamma}_k$ at any finite set of distinct locations is nonsingular \citep{chang1996strictly, wendland2005scattered}. However, the eigenvalues of the integral operator defined by an analytic kernel decay geometrically \citep{little1984eigenvalues}, and the covariance matrix inherits this decay as the locations fill a fixed interval, so under in-fill asymptotics the smallest eigenvalue approaches zero and the condition number grows rapidly.

The target PSD, in contrast, need not be compactly supported. In continuous time or space it may be positive at arbitrarily high frequencies. In Section~\ref{sec:assumptions} this is represented via the truncation error $\varepsilon_a$, which is made arbitrarily small in Theorem~\ref{the:jackson} by widening the knot span. Although this suffices in theory, in practice widening at a fixed knot spacing adds additional parameters to estimate. A natural feature of applied problems limits this: a physical process commonly has some notion of smallest scale of variation \citep[e.g. the scale set by viscous dissipation in a turbulent flow, ][Chapter~6]{pope2000turbulent}. Beyond the corresponding frequency the spectral density is negligible and we may take $\varepsilon_a \approx 0$. Variation in the data above such frequency is typically noise, and may be modelled apart from the spline expansion. For instance, \citet{im2007semiparametric} join a cubic B-spline expansion of an isotropic spectral density to a truncated algebraic tail above a cutoff frequency.

A viable alternative is to assume the data instead lie on a (potentially gappy) lattice of spacing $\Delta$ \citep[for inference on incomplete lattices see][]{stroud2017bayesian, guinness2017circulant}. Suppose the aliased spectral density of the sampled process lies between $m_\Delta > 0$ and $M_\Delta < \infty$ on the Nyquist band $B_\Delta = [-1/(2\Delta), 1/(2\Delta)]$; then every eigenvalue of the covariance matrix lies in $[m_\Delta/\Delta, M_\Delta/\Delta]$, and the condition number is at most $M_\Delta/m_\Delta$, uniformly in the number of observations and the pattern of gaps \citep[Section 4.2, Lemma 4.1]{gray2006toeplitz}. Further, confining the knots to the Nyquist band removes the truncation term from Theorem~\ref{the:jackson}, and Corollary~\ref{cor:lattice_jackson} states the approximation error of the corresponding auto-covariance sequence. Although theoretically pleasing, such an assumption on the data is a user-decision and should not be naively stipulated.

\begin{corollary} \label{cor:lattice_jackson}
  Let $f \in \mathcal{W}^{r,p}(B_\Delta)$ for some $1 \leq p \leq 2$ with $f$ supported in $B_\Delta$, and let the knot sequence $\mathcal{K}$ partition $B_\Delta$ subject to the conditions of Theorem~\ref{the:jackson}. Then there exists $\hat{f}_k \in \mathcal{S}_{k, \mathcal{K}}$ such that, for $1/p + 1/q = 1$,
  \begin{equation*}
    \Delta^{1/q} \norm{\gamma - \hat{\gamma}_k}_{\ell^q(\Delta \mathbb{Z})} \leq \norm{f - \hat{f}_k}_{L^p(B_\Delta)} \leq C_{k,p} h_{\max}^{r} \norm{f^{(r)}}_{L^p(B_\Delta)}.
  \end{equation*}
\end{corollary}

\section{Multivariate Random Processes} \label{sec:multivariate}

Assume now that $X(t)$ is multivariate such that $M > 1$. Cram\'{e}r's Representation Theorem extends the conditions of Bochner's Theorem in \eqref{eqn:bochner} and states that a family $\{\gamma^{(r,s)}\}_{r,s=1}^M$ of integrable functions is the ACF of a weakly-stationary multivariate stochastic process if and only if they admit the representation
\begin{equation*}
  \gamma^{(r,s)}(\tau) = \int_\mathbb{R} f^{(r,s)}(\omega) e^{2 \pi \iota \omega \tau} \; \rd \omega, \quad \text{for all} \quad r,s \in \{1, \dots, M\}
\end{equation*}
for a Hermitian, matrix-valued function $\bbf(\omega) = \{f^{(r,s)}(\omega)\}_{r,s=1}^M$ with integrable entries such that $\bbf(\omega)$ is pointwise positive semi-definite almost everywhere. The principal diagonal of $\bbf(\omega)$ thus is required to be real-valued and non-negative; however, the off-diagonal terms may be complex-valued as long as $\bbf(\omega)$ respects positive semi-definiteness.

Extension of the univariate spline kernel in Section~\ref{sec:univariate} is simple, whereby we model each of the
\begin{equation} \label{eqn:mulivariate_spline}
  \hat{f}^{(r,s)}_k(\omega) = \sum_{i=0}^{m_{r,s}-1} c_i^{(r,s)} B_{i,k}(\omega; \mathcal{K}_{r,s}) \quad \Longrightarrow \quad \hat{\gamma}^{(r,s)}_k(\tau) = \sum_{i=0}^{m_{r,s}-1} c_i^{(r,s)} \rho_{i,k}(\tau; \mathcal{K}_{r,s})
\end{equation}
where $c_i^{(r,s)} = a_i^{(r,s)} + \iota b_i^{(r,s)}$ is complex-valued and  describes the $i$th coefficient for the $(r,s)$th cross-spectral density, and the spline and ACF bases $B_{i,k}(\omega; \mathcal{K}_{r,s})$ and $\rho_{i,k}(\tau; \mathcal{K}_{r,s})$ are defined with an $(r,s)$-specific knot placement $\mathcal{K}_{r,s}$. Each cross-spectral density need not share the same number of bases, and so the summation is with respect to $m_{r,s}$. Hermitian symmetry is enforced by setting $\mathcal{K}_{r,s} = \mathcal{K}_{s, r}$ and $c_i^{(r,s)} = \overline{c_i^{(s, r)}}$. Cram\'{e}r's Theorem imposes constraints on the coefficients $c_i^{(r,s)}$ to ensure that $\bbf(\omega)$ remains positive semi-definite for all $\omega$. Expressing \eqref{eqn:mulivariate_spline} on the union $\mathcal{K} = \bigcup_{r,s} \mathcal{K}_{r,s}$,
\begin{equation*}
  \hat{f}_k^{(r,s)}(\omega) = \sum_{i=0}^{m-1} c_i^{(r,s)} B_{i,k}(\omega; \mathcal{K}),
\end{equation*}
as $B_{i,k}(\omega; \mathcal{K}) \geq 0$, and as positive semi-definite matrices are closed under non-negative linear combinations, a sufficient condition that satisfies Cram\'{e}r's Theorem is if all $i$th coefficient matrices $\textbf{C}_i = \{c_i^{(r,s)}\}_{r,s=1}^M$ are Hermitian and positive semi-definite.  This condition is sufficient but not necessary, see Section~\ref{sec:nonnegativity} for further discussion.

As discussed in the introduction, two analogous spectral kernel constructions have been proposed for modelling multivariate stationary processes: the multivariate spectral mixture kernel \citep{parra2017spectral} and the Minecraft kernel \citep{simpson2021minecraft}. However, as noted in \cite{simpson2021minecraft}, the spectral mixture kernel cannot represent the full range of cross-spectra allowed by Cram\'{e}r's Theorem, as its cross-spectra are built from overlapping Gaussian components. In contrast, our construction places a common B-spline basis under every entry of $\bbf$, with a positive semi-definite coefficient matrix $\textbf{C}_i$ for each basis function. The following result establishes that these B-spline-based kernels are dense in the space of multivariate stationary covariance functions.

\begin{theorem} \label{the:multivariate}
  Denote by $\Lambda_M$ the set of weakly stationary, $M \times M$ matrix-valued covariance functions $\bgamma : \mathbb{R} \to \mathbb{C}^{M \times M}$ that admit a spectral density $\bbf$, which is Hermitian and positive semi-definite for almost every $\omega$ with entries $f_{ij} \in L^1(\mathbb{R})$. Let $\hat{\Lambda}_{M,k}$ denote the set of inverse Fourier transforms $\hat{\bgamma}_k$ of $\hat{\bbf}_k = \sum_{i=0}^{m-1} \textbf{C}_i B_{i,k}(\cdot\,; \mathcal{K})$, for some knot sequence $\mathcal{K}$ and Hermitian positive semi-definite $\textbf{C}_i \in \mathbb{C}^{M \times M}$. Then $\hat{\Lambda}_{M,k} \subset \Lambda_M$ and, for any $\bgamma \in \Lambda_M$ and any $\varepsilon_\gamma > 0$, there exists $\hat{\bgamma}_k \in \hat{\Lambda}_{M,k}$ such that $\sum_{i,j=1}^M \| f_{ij} - \hat{f}_{ij,k} \|_{L^1(\mathbb{R})} < \varepsilon_\gamma$, and hence $\sum_{i,j=1}^M \sup_{\tau \in \mathbb{R}} | \gamma_{ij}(\tau) - \hat{\gamma}_{ij,k}(\tau) | < \varepsilon_\gamma$. So $\hat{\Lambda}_{M,k}$ is dense in $\Lambda_M$ under the entrywise $L^1$ norm on spectral densities, and hence under the entrywise supremum norm on ACFs.
\end{theorem}

Known or assumed structure in $\gamma^{(r,s)}_k(\tau)$ can reduce the number of coefficients required for estimation \citep[see][]{parra2017spectral}. If $\gamma^{(r,s)}(\tau)$ is assumed to be real-valued and symmetric in $\tau$, then the corresponding cross-spectral density $f^{(r,s)}(\omega)$ must be real and even. In this case, it suffices to use the symmetrised basis $\frac{1}{2}[B_{i,k}(\omega) + B_{i,k}(-\omega)]$ of Section~\ref{sec:univariate} and restrict the coefficients to be real-valued $c_i^{(r,s)} = a_i^{(r,s)} \in \mathbb{R}$. This assumption precludes the possibility of modeling lead--lag structure between components, but halves the number of free parameters. Next, assume the cross-covariance function is a time-delayed version of a symmetric ACF, $ \hat{\gamma}^{(r,s)}_k(\tau) = \tilde{\gamma}^{(r,s)}_k(\tau - t_0)$, for some fixed delay $t_0$ that is shared across all frequency components. This corresponds to multiplying the spectral representation by a global linear phase term 
\[\hat{f}^{(r,s)}_k(\omega) = \tilde{f}^{(r,s)}_k(\omega) e^{-2\pi i \omega t_0} = e^{-2\pi i \omega t_0} \sum_{i=0}^{m_{r,s}-1} a_i^{(r,s)} B_{i,k}(\omega; \mathcal{K}_{r,s}) .\]
This phase term preserves the amplitude spectrum $|\hat{f}^{(r,s)}_k(\omega)| = |\tilde{f}^{(r,s)}_k(\omega)|$. Therefore, if the ACF is known to be real, symmetric, and simply delayed, the amplitude spectrum alone suffices to model its shape and the delay can be introduced separately as a global phase modulation. This corresponds to a special case of \eqref{eqn:mulivariate_spline} where the phase structure is globally shared across the basis. Even if not explicitly assumed, such structures may emerge as parsimonious representations during inference.

\section{Multidimensional Random Processes} \label{sec:multidimensional}

A natural extension of the univariate B-spline construction over multidimensional domains (e.g. space or space-time) is to consider tensor-product B-splines. For simplicity, throughout this section we fix $M=1$; the case of $M,D>1$ follows by combining Sections~\ref{sec:multivariate} and \ref{sec:multidimensional}. Suppose we are interested in modelling a PSD defined over a product domain $\boldsymbol{\omega} = (\omega_1, \dots, \omega_D) \in \mathbb{R}^D$. For each coordinate direction $\omega_j$, we specify a univariate B-spline basis $\{B^{(j)}_{i_j, k_j}(\omega_j)\}_{i_j = 0}^{m_j - 1}$, where the degree $k_j$ and knot set $\mathcal{K}^{(j)} = \{\kappa^{(j)}_0 < \dots < \kappa^{(j)}_{m_j+k_j}\}$ are defined analogously to the univariate case. A diagram of the gridded knot layout is shown in the left panel of Figure~\ref{fig:mesh}, in Appendix~\ref{app:adaptivity}. The full tensor-product basis over $\mathbb{R}^D$ is then constructed as
\[
B_{\mathbf{i}, \mathbf{k}}(\boldsymbol{\omega}) = \prod_{j=1}^D B^{(j)}_{i_j, k_j}(\omega_j), \quad \mathbf{i} = (i_1, \dots, i_D),
\]
which defines a basis for the multidimensional spline space $\mathcal{S}_{\mathbf{k}, \boldsymbol{\mathcal{K}}} = \mathrm{span}\{B_{\mathbf{i}, \mathbf{k}}\}$. 

Extending \eqref{eqn:psd_basis} to the multidimensional setting, we approximate the PSD as
\[
\hat{f}_{\mathbf{k}}(\boldsymbol{\omega}) = \sum_{\mathbf{i}} c_{\mathbf{i}} B_{\mathbf{i}, \mathbf{k}}(\boldsymbol{\omega}),
\]
with coefficients $c_{\mathbf{i}}$ ensuring $\hat{f}_{\mathbf{k}}(\boldsymbol{\omega}) \geq 0$ for all $\boldsymbol{\omega} \in \mathbb{R}^D$. As in the univariate case, $c_{\mathbf{i}} \geq 0$ is sufficient, since a tensor-product B-spline is a product of non-negative functions. Further, as tensor-product B-splines factor as products of univariate bases, their inverse Fourier transforms also factor as
\begin{equation} \label{eqn:multivariate_rho}
  \rho_{\mathbf{i}, \mathbf{k}}(\boldsymbol{\tau}) = \prod_{j=1}^D \rho^{(j)}_{i_j, k_j}(\tau_j),
\end{equation}
where each $\rho^{(j)}_{i_j, k_j}(\tau_j)$ is the autocovariance basis corresponding to $B^{(j)}_{i_j, k_j}(\omega_j)$, as given in Theorem~\ref{the:acf_closed_form}. The multidimensional autocovariance function is then expressed as
\begin{equation} \label{eqn:multidimensional}
\hat{\gamma}_{\mathbf{k}}(\boldsymbol{\tau}) = \sum_{\mathbf{i}} c_{\mathbf{i}} \rho_{\mathbf{i}, \mathbf{k}}(\boldsymbol{\tau}) = \sum_{\mathbf{i}} c_{\mathbf{i}} \prod_{j=1}^D \rho^{(j)}_{i_j, k_j}(\tau_j).
\end{equation}
Although each component $\rho_{\mathbf{i}, \mathbf{k}}(\boldsymbol{\tau})$ is separable in the dimensions of $\boldsymbol{\tau}$, the sum need not be. Equation~\eqref{eqn:multidimensional} is separable if and only if $c_{\mathbf{i}} = \prod_j c^{(j)}_{i_j}$, in which case the sum factorises as $\prod_j \sum_{i_j} c^{(j)}_{i_j} \rho^{(j)}_{i_j, k_j}(\tau_j)$, a product of univariate functions. Separability is therefore the exceptional case; the required symmetrisation forces $\hat\gamma_{\mathbf k}$ to be even in each coordinate, which excludes propagating structure. Jackson-type convergence results analogous to Theorem~\ref{the:jackson} hold in the multidimensional case, with convergence rates governed by the spline degrees and knot resolution along each axis.

Tensor-product bases have two limitations. They are not locally adaptive, as inserting a single knot in the $j$th direction introduces $\prod_{l \neq j} m_l$ new basis functions; and every basis function is supported on a box aligned with the coordinate axes, so the class is anisotropic only along those axes. Appendix~\ref{app:adaptivity} addresses these constraints: the hierarchical and T-mesh constructions of Appendices~\ref{sec:hierarchical} and~\ref{sec:t_mesh} allow local refinement, and Appendix~\ref{sec:rotation} evaluates the basis at a linearly transformed frequency, which removes the axis alignment and keeps the ACF in closed form.

\section{Inference in practice} \label{sec:inference}

Sections~\ref{sec:univariate}--\ref{sec:multidimensional} provide a description of the ACF and are agnostic to the inference scheme. For instance, coefficients may be estimated by likelihood methods, or assigned a prior and inferred within a Bayesian framework; we use maximum likelihood throughout the remaining sections. This section discusses practical inference concerns: Section~\ref{sec:likelihood} documents two viable likelihoods that may be employed and Section~\ref{sec:nonnegativity} addresses inference methodologies that respect the non-negativity constraints assumed throughout.

\subsection{Choice of likelihood} \label{sec:likelihood}

When estimating the second-order properties of a random process $X(t)$ via likelihood-based methodologies we typically use either the Gaussian or the Whittle likelihoods. These are standard likelihoods for when $X(t)$ is Gaussian, but as we will discuss, both are suitable quasi-likelihoods for non-Gaussian $X(t)$. The covariance model and the spectral model are related by the Fourier transform, and so the choice of likelihood is dependent on the structure of the data and the available compute. Denote by $\textbf{x} = (x_1, \ldots, x_n)^\T$ the observations of the process at locations $\bt_1, \ldots, \bt_n$, and denote by $\bm{\Sigma}$ the covariance matrix with $i,j$-th entries $\bm{\Sigma}_{ij} = \hat{\gamma}_k(\bt_i - \bt_j)$, formed in closed form via Theorem~\ref{the:acf_closed_form}. Strictly, the covariance matrix $\bm{\Sigma}$ is a function of the coefficients $\{c_i\}$ although we suppress this dependency in the notation. The exact Gaussian log-likelihood is
\begin{equation} \label{eqn:gaussian_lik}
  \mathcal{L}_\mathrm{G}(c) = - \tfrac{1}{2} \log \lvert \bm{\Sigma} \rvert - \tfrac{1}{2} \textbf{x}^\T \bm{\Sigma}^{-1} \textbf{x},
\end{equation}
which is available for any configuration of observation locations. The naive cost is $\mathcal{O}(n^3)$ operations and $\mathcal{O}(n^2)$ storage, for a dense Cholesky factorisation. Whittle's likelihood provides a computationally efficient, frequency domain approximation to \eqref{eqn:gaussian_lik}. Up to an additive constant, the Whittle log-likelihood is
\begin{equation} \label{eqn:whittle}
  \mathcal{L}_\mathrm{W}(c) = - \sum_{l \in \Omega_n} \left\{ \log \hat{f}_k(\bomega_l) + \frac{I(\bomega_l)}{\hat{f}_k(\bomega_l)} \right\},
\end{equation}
where $\Omega_n$ is the set of Fourier frequencies, $\hat{f}_k$ denotes the model of the spectral density evaluated at discrete Fourier frequencies $\bomega_l$, and $I(\bomega_l)$ is the periodogram of the data. Compared with the Gaussian likelihood, \eqref{eqn:whittle} only requires $\mathcal{O}(n \log n)$ operations; however, it further requires observations on a regular grid so that the discrete Fourier transform to compute $I(\bomega_l)$ is available. Substituting the spline model $\hat{f}_k(\bomega) = \sum_i c_i B_{i,k}(\bomega)$, the gradient and Hessian of \eqref{eqn:whittle} with respect to the coefficients $\{c_i\}$ are respectively given by
\begin{align*}
  \frac{\partial \mathcal{L}_\mathrm{W}}{\partial c_i} = \sum_{l \in \Omega_n} \left\{ \frac{I(\bomega_l)}{[\hat{f}_k(\bomega_l)]^2} - \frac{1}{\hat{f}_k(\bomega_l)} \right\} B_{i,k}(\bomega_l), \quad
  \frac{\partial^2 \mathcal{L}_\mathrm{W}}{\partial c_i \partial c_j} = \sum_{l \in \Omega_n} \left\{ \frac{1}{[\hat{f}_k(\bomega_l)]^2} - \frac{2 I(\bomega_l)}{[\hat{f}_k(\bomega_l)]^3} \right\} B_{i,k}(\bomega_l) B_{j,k}(\bomega_l).
\end{align*}
Since each $B_{i,k}(\bomega)$ has compact support, the gradient with respect to $c_i$ receives contributions from only a small number of frequencies, and the Hessian entry $\partial^2 \mathcal{L}_\mathrm{W} / \partial c_i \partial c_j$ is non-zero only when the supports of $B_{i,k}$ and $B_{j,k}$ overlap. The Hessian is therefore sparse, with at most $(2k+1)^D$ non-zero entries in each row, and the localised influence of each parameter simplifies the optimisation landscape. Although the objective \eqref{eqn:whittle} is non-convex in general and thus does not guarantee global convergence, the sparsity and local structure substantially reduce the risk of pathological optimisation behaviour and improve numerical stability in practice. When $X(t)$ is non-Gaussian, in the absence of a distributional assumption that admits an exact likelihood, both \eqref{eqn:gaussian_lik} and \eqref{eqn:whittle} offer a suitable surrogate. Both are Gaussian quasi-likelihoods whose estimating equations only depend on the data through second moments, and each is consistent for the second-order structure \citep{guyon1982parameter}. For instance, the Whittle estimator remains consistent without Gaussianity, and departures from Gaussianity enter only through its higher-order cumulant spectra. Note that, without Gaussianity, \eqref{eqn:gaussian_lik} and \eqref{eqn:whittle} cannot guarantee the standard efficiency and standard errors, and so the fit coefficients should be interpreted as a second-order summary of the field. Interval estimates may be available with further assumptions and corrections \citep[e.g.][]{kreiss2023bootstrapping}.

\subsection{Non-negativity of the fit spectral density} \label{sec:nonnegativity}

Equation~\eqref{eqn:psd_basis}, and the multivariate and multidimensional extensions, require the coefficients to be such that the fit spectral density is non-negative. As the B-splines are themselves non-negative at every degree and for every knot sequence \citep[Chapter~4]{schumaker2007spline}, it is sufficient to enforce $c_i \geq 0$ for all $i$. For $k \leq 1$ the coefficients equate to values taken by $\hat{f}_k$, on each knot interval for $k = 0$ and at the knots for $k = 1$, and $c_i \geq 0$ is necessary as well as sufficient. For $k \geq 2$, the assumption $c_i \geq 0$ is strictly stronger than non-negativity of $\hat{f}_k$ \citep{burchard1973extremal}; \citet{de1974splines} show that a non-negative function may be approximated by splines with non-negative coefficients to $\mathcal{O}(h_{\max}^2)$, and no better in general. Nonetheless, the exact feasible set is convex as $\hat{f}_k$ is linear in $c$. Writing $b(\omega) = (B_{0,k}(\omega), \ldots, B_{m-1,k}(\omega))^\T$, the set $\{ c \in \mathbb{R}^m : b(\omega)^\T c \geq 0 \text{ for all } \omega \}$ is an intersection of half-spaces through the origin and hence a closed convex cone. Strictly, this cone is defined by infinitely many linear constraints, corresponding to each value of $\omega$, although these are equivalent to finitely many semidefinite constraints. Following \citet{papp2014shape}, as $\hat{f}_k$ is a polynomial of degree $k$ on each knot interval, non-negativity on each interval can be ensured by a semidefinite program of size linear in $m$. Both the multivariate and multidimensional extensions inherit the sufficient condition: tensor-product B-splines are products of non-negative functions, and non-negatively weighted sums of Hermitian positive semi-definite matrices are again Hermitian positive semi-definite, so it suffices to take $c_{\mathbf{i}} \geq 0$ in the multidimensional model and $\textbf{C}_i \succeq 0$ in the multivariate model. We expect the conditions of \citet{papp2014shape} to extend to the multivariate case, since for $M > 1$ the spectral density matrix is, on each knot interval, a matrix-valued polynomial in a single variable. However, for $D > 1$ the splines are piecewise polynomials in several variables, for which non-negativity no longer reduces to a single semidefinite constraint; exact verification instead requires a hierarchy of semidefinite relaxations \citep{lasserre2001global}. The exception is $k_j \leq 1$ in every direction, where $c_{\mathbf{i}} \geq 0$ is again necessary and sufficient. It is important to note that characterisations of this kind are most useful when the objective is convex in $c$, as for the least-squares criteria of \citet{papp2014shape}, as the whole fit is then a single convex program. This is not the case for either \eqref{eqn:gaussian_lik} or \eqref{eqn:whittle} in Section~\ref{sec:likelihood}, so imposing the exact constraint would require a sequence of semidefinite programs rather than one. We instead impose $c_i \geq 0$ throughout. For $k \leq 1$ this is exactly non-negativity of $\hat{f}_k$; for $k \geq 2$ it can reduce the rate of Theorem~\ref{the:jackson} to $\mathcal{O}(h_{\max}^2)$ in the worst case \citep{de1974splines}. 

\section{Simulation Studies} \label{sec:simulation}

We present two simulation studies that study repeated samples of GPs. First, we study the performance under different choices of $k$ and $\mathcal{K}$, for a Mat\'{e}rn-$3/2$ model with exponential tail decay and a Cauchy ACF with power-law decay, both with varying length scales. Next, we look at the non-separable bivariate Mat\'{e}rn of \citet{gneiting2010matern} across differing degrees of cross-correlation. In all simulations we estimate coefficients via maximum likelihood using the full Gaussian likelihood.

\subsection{Univariate Simulation} \label{sec:uni_sim}

We simulate GP draws sampled at interval $ \Delta = 1 $ from two ACFs: a Matérn-$3/2$ ACF with exponentially decaying tail, $\gamma(\tau) = \sigma^2 (1 + \sqrt{3} \lvert \tau \rvert/\ell ) \exp(-\sqrt{3} \lvert \tau \rvert/\ell)$, and a Cauchy ACF, $\gamma(\tau) = \sigma^2 (1 + \lvert \tau \rvert/\ell)^{-1}$, whose tail decays as $\lvert \tau \rvert^{-1}$ and is not integrable, so that its spectral density is unbounded at the origin. We set sample lengths of $n = 2000$ for both ACFs. As the Cauchy spectral density has a singularity at the origin, each Cauchy record is centred by its sample mean before fitting. We fix the variance at $ \sigma^2 = 1 $ and consider length-scale values $ \ell \in \{2, 5, 10, 20\} $. For each combination of spline degree $ k \in \{0, 1, 2\} $ and number of knots $ n_{\mathrm{knots}} \in \{2, 4, 8, 16\} $ for the Matérn ACF, and $ n_{\mathrm{knots}} \in \{4, 8, 16\} $ for the Cauchy ACF, we perform 1000 independent simulations. For the Matérn ACF, knots are placed equally in an offset-log domain, i.e., equally spaced in $ \log(\omega + b) $ with offset $ b = 0.01 $, over the interval $ [0, 0.5] $, bounded within the Nyquist frequency. This yields approximately log-spaced knots while preserving resolution near the origin. For the Cauchy ACF, knots are placed so that each basis support contains at least six Fourier frequencies. For $ k > 0 $, we extend this knot sequence by $ k $ locations beyond each boundary to accommodate the spline basis support. As a benchmark, we compute the unbiased empirical ACF,
\begin{equation} \label{eqn:empirical}
\hat{\gamma}_{\mathrm{emp}}(\tau) = \frac{1}{n - \tau} \sum_{t=1}^{n - \tau} x_t x_{t+\tau},
\end{equation}
noting that this estimator is not guaranteed to yield a positive semi definite estimate. We report the integrated absolute error (IAE) between the true and estimated ACF,
\begin{equation} \label{eqn:IAE}
\mathrm{IAE} = \frac{1}{10\ell} \int_0^{10\ell} \left| \gamma(\tau) - \hat{\gamma}_{k,\mathcal{K}}(\tau) \right| \, \mathrm{d}\tau,
\end{equation}
as a summary of the error. The upper bound $ 10\ell$ ensures comparability across different length-scales. The integral is approximated as a Riemann sum. The empirical ACF is defined only at the integer lags, so its IAE is a sum of absolute errors over the lattice lags. This aligns with the view given in Section~\ref{sec:continuous_domain} where data are considered to live on a lattice of spacing $\Delta = 1$.

\begin{figure}[pos=H]
  \centering
  \includegraphics[width=0.8\textwidth]{"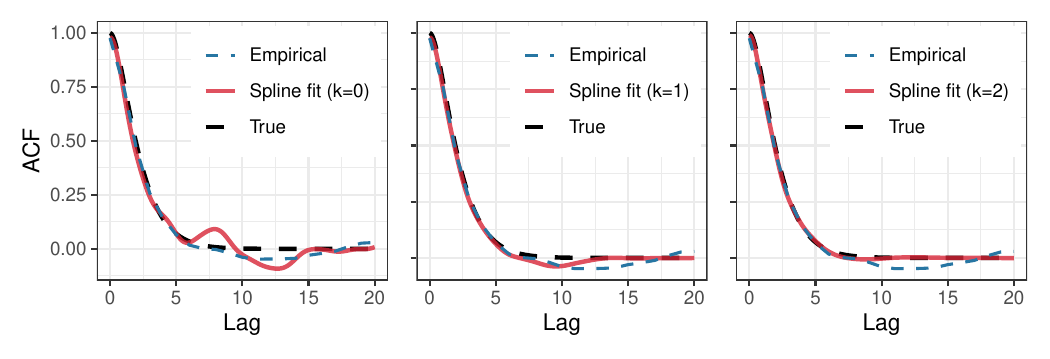"}
  \caption{Estimated ACFs for a GP with a Matérn-$3/2$ ACF, based on a single simulated realisation with $\ell = 2$ and $n_{\mathrm{knots}} = 4$. The true ACF is shown as a thick black dashed line, and the empirical ACF estimate is shown as the thin blue dashed line. The three panels correspond to spline-based estimates with $k = 0$ (left), $k = 1$ (middle), 
  and $k = 2$ (right), each plotted in red.
  }
  \label{fig:1d_acfs}
\end{figure}

We show a single fit from the Mat\'ern-3/2 model with $\ell = 2$ and $n_{\mathrm{knots}} = 4$ in Figure~\ref{fig:1d_acfs}. The true ACF is given by the thick black dashed line, and the empirical estimate by the thin blue dashed line; these values are the same in each plot. The three panels correspond to the $k=0$ (left), $k=1$ (middle) and $k=2$ (right) fits, respectively, depicted in red. The three fits in Figure~\ref{fig:1d_acfs} differ most in the tails. Corollary~\ref{cor:acf_tail_bound} provides rationale for this behaviour: for $k \geq 1$ the tail error decays as $\mathcal{O}(|\tau|^{-k})$, whereas for $k = 0$ the corollary admits only $j = 0$ and gives no decay in $\tau$. This is seen in the increasing damping of the tail behaviour as $k$ increases. Table~1 reports the mean and standard deviation of the IAEs between the true and modelled ACFs, across the 1000 simulations and scaled by $ \times 10^2 $, with the Matérn ACF in the left half and the Cauchy ACF in the right. For the Matérn ACF, the spline-based estimator consistently outperforms the empirical ACF across most configurations, with the exception of spline degree $ k = 0$ \citep[corresponding to the model of][]{tobar2019band} and cases with only $ n_{\mathrm{knots}} = 2 $. The case of $k=0$ empirically demonstrates our claim that convergence can be faster for $k > 0$, and the low-knot regimes correspond to under-parameterised models that struggle to capture the model complexity. Notably, in general we provide more accurate estimates of the true ACF whilst ensuring positive semi-definiteness, a property not guaranteed by the empirical estimator. While increasing $n_{\mathrm{knots}}$ improves flexibility, it may also incur a small penalty due to the bias-variance trade-off. Estimation is most accurate when the length-scale $ \ell $ is small for two reasons: the ACF steepens, increasing the effective sample size of the data; and the Matérn-$3/2$ spectrum exhibits a broad low-frequency plateau that aligns well with the smaller basis representations. For the Cauchy ACF, the spline estimator with $k \geq 1$ matches the empirical ACF to within $0.3$ in every configuration. Here the error is dominated by the few lowest Fourier frequencies, near the singularity at the origin, where the periodogram offers too few ordinates for smoothing to reduce the variance. Overall, the results demonstrate that moderate spline degree ($ k = 1 $ or $ 2 $) combined with a sufficient number of knots provides robust and accurate recovery of the ACF, with the spline-based estimator matching or improving on the empirical estimate, whilst guaranteeing positive semi-definiteness.

\renewcommand{\arraystretch}{0.9}
\begin{table}[t!]
\begin{minipage}{\textwidth}
\caption{Integrated absolute error between the true and estimated ACF, scaled by $ \times 10^2 $, for Gaussian processes with, left, a Matérn-$3/2$ ACF, and, right, a Cauchy ACF $\gamma(\tau) = (1 + \lvert \tau \rvert / \ell)^{-1}$, both of length $n = 2000$ at unit sampling interval. Each entry reports the mean IAE over 1000 simulations for a given combination of spline degree $k$, number of interior knots $n_{\mathrm{knots}}$, and length-scale $\ell$. Standard deviations are shown in parentheses. The ACF is estimated using the spline-based method described in Theorem~\ref{the:acf_closed_form} in the main text, and results are compared with the unbiased empirical ACF: for the Matérn ACF, bold values achieve a better IAE than the empirical ACF; for the Cauchy ACF, bolding is omitted as the spline estimates with $k \geq 1$ lie within $0.3$ of the empirical ACF. All errors are computed over the interval $[0, 10\ell]$ as a proxy for $L^1$ error. For the Cauchy ACF, 35 of the 36000 spline fits failed to converge and are omitted from their cells.}
\centering
{\small\setlength{\tabcolsep}{3pt}
\begin{tabular}{cc | cccc | ccc}
  \toprule
& & \multicolumn{4}{c|}{Mat\'{e}rn-$3/2$} & \multicolumn{3}{c}{Cauchy} \\
& & \multicolumn{4}{c|}{$n_\mathrm{knots}$} & \multicolumn{3}{c}{$n_\mathrm{knots}$} \\
$\ell$ & $k$ & $2$ & $4$ & $8$ & $16$ & $4$ & $8$ & $16$ \\
\midrule
\multirow{4}{*}{2}
  & 0 & 7.7 (0.13) & 3.8 (0.64) & \textbf{2.7 (0.81)} & \textbf{2.8 (1.1)} & 6.1 (0.83) & 3.3 (1.7) & 3.1 (1.8) \\
  & 1 & \textbf{2.7 (0.74)} & \textbf{2.1 (0.95)} & \textbf{3 (1.2)} & \textbf{3.2 (1.1)} & 3 (1.7) & 3 (1.9) & 3.1 (1.9) \\
  & 2 & \textbf{1.1 (0.6)} & \textbf{1.7 (0.74)} & \textbf{2.8 (1)} & 3.3 (1.3) & 3.1 (1.9) & 3.1 (1.9) & 3.1 (1.9) \\
  & $\mathrm{emp.}$ & 3.2 (1) & 3.2 (1) & 3.2 (1) & 3.2 (1) & 3.1 (1.8) & 3.1 (1.8) & 3.1 (1.8) \\
\midrule
\multirow{4}{*}{5}
  & 0 & 9.7 (0.11) & 7.6 (0.27) & \textbf{2.7 (0.81)} & \textbf{3.7 (1.5)} & 6.6 (1.9) & 5 (2.8) & 4.9 (3) \\
  & 1 & \textbf{2.5 (0.52)} & \textbf{2.6 (1.2)} & \textbf{3 (1.2)} & \textbf{4.6 (1.7)} & 4.9 (3) & 4.8 (3.1) & 5 (3.1) \\
  & 2 & 5.8 (0.55) & \textbf{2 (0.57)} & \textbf{2.8 (1)} & \textbf{3.7 (1.8)} & 4.8 (3.1) & 5 (3.3) & 5 (3.1) \\
  & $\mathrm{emp.}$ & 5.1 (1.9) & 5 (1.9) & 5 (1.9) & 5 (1.9) & 4.9 (2.9) & 4.9 (2.9) & 4.9 (2.9) \\
\midrule
\multirow{4}{*}{10}
  & 0 & 10 (0.11) & 9.3 (0.19) & \textbf{7.2 (0.55)} & \textbf{4.4 (1.5)} & 8.9 (3.2) & 6.6 (3.5) & 6.6 (3.6) \\
  & 1 & 7.3 (0.43) & \textbf{3.1 (0.76)} & \textbf{2.5 (1.1)} & \textbf{ 4.1 (2.7)} & 6.5 (3.6) & 6.7 (3.8) & 6.7 (3.8) \\
  & 2 & 9.4 (0.45) & \textbf{6.7 (0.55)} & \textbf{2.7 (0.74)} & \textbf{ 2.5 (1.3)} & 6.4 (3.9) & 6.5 (3.7) & 6.7 (3.8) \\
  & $\mathrm{emp.}$ & 7.2 (2.9) & 7.2 (2.9) & 7.2 (2.9) & 7.2 (2.9) & 6.7 (3.5) & 6.7 (3.5) & 6.7 (3.5) \\
\midrule
\multirow{4}{*}{20}
  & 0 & 11 (0.1) & \textbf{10 (0.27)} & \textbf{9.3 (0.52)} & \textbf{7.4 (1.1)} & 9.6 (4.4) & 9.3 (4.9) & 9.1 (4.8) \\
  & 1 & \textbf{9.9 (0.35)} & \textbf{7.6 (0.6)} & \textbf{3.3 (0.98)} & \textbf{3.5 (2)} & 9.5 (5.5) & 9.5 (5.3) & 9.5 (5.3) \\
  & 2 & 11 (0.19) & \textbf{8.6 (0.55)} & \textbf{7.2 (0.61)} & \textbf{3.2 (1.2)} & 9.5 (5.3) & 9.4 (5.6) & 9.3 (5.2) \\
  & $\mathrm{emp.}$ & 10 (4.4) & 10 (4.4) & 10 (4.4) & 10 (4.4) & 9.3 (5.1) & 9.3 (5.1) & 9.3 (5.1) \\
  \bottomrule
\end{tabular}}
\end{minipage}
\label{tab:uni_sim_results}
\end{table}
\renewcommand{\arraystretch}{1}

\subsection{Multivariate Simulation}

Next, we extend the simulation study to the non-separable bivariate Mat\'{e}rn ACF of \cite{gneiting2010matern}. Define $\gamma_{11}$ and $\gamma_{22}$ as the true marginal ACFs, and $\gamma_{12}$ as the cross-covariance function, all with length-scale $\ell = 2$. We set $\gamma_{11}$ and $\gamma_{22}$ to be Mat\'{e}rn ACFs with roughness parameters $\nu_{11} = 2$ and $\nu_{22} = 1$, respectively, and variance $\sigma_{11}^2 = \sigma_{22}^2 = 1$. The cross-covariance function is similarly specified as a Mat\'{e}rn function with $\nu_{12} = 1.5$, with a correlation term $\lambda_{12} \in \{-0.9, -0.5, 0, 0.5, 0.9\}$ that we vary over the simulation.  We present results for $k \in \{0, 1, 2\}$, with $3$ fixed interior knots evenly spaced with offset-log spacing as in Section~\ref{sec:uni_sim}. Each replicate is a bivariate record of length $n = 1000$ at unit sampling interval. Table~2 reports the results for $\gamma_{11}$ and $\gamma_{12}$; the $\gamma_{22}$ results are similar to $\gamma_{11}$ and omitted for brevity. Reported values represent the mean IAEs and standard deviations, calculated over a 1000 member ensemble and as per \eqref{eqn:IAE}. Our methodology is compared to the empirical estimator in \eqref{eqn:empirical}, and the fit ACF with the knowledge of the true parametric form, calculated via maximum likelihood. Such perfect knowledge is rare in practice and it is included here as a notion of a viable lower bound. Finally, recall that the case of $k=0$ corresponds to the model of \cite{tobar2019band}.

\renewcommand{\arraystretch}{0.9}
\begin{table}
\begin{minipage}{\textwidth}
\caption{
Mean integrated absolute errors from 1000 replications of marginal and cross-covariance functions $\gamma_{11}$ and $\gamma_{12}$ from a bivariate Matérn process with varying cross-correlation $\lambda_{12} \in \{-0.9, -0.5, 0, 0.5, 0.9\}$, each replicate of length $n = 1000$ at unit sampling interval. Standard deviations are in parentheses. Reported estimators include spline-based models with B-spline degree $k \in \{0, 1, 2\}$, the empirical ACF estimator, and the parametric maximum likelihood estimator with access to the true model class. Bold values achieve a better IAE than the empirical ACF.}
\centering
\begin{tabular}{cc | ccccc}
  \toprule
& & \multicolumn{5}{c}{$\lambda_{12}$} \\
 & $k$ & $-0.9$ & $-0.5$ & $0$ & $0.5$ & 0.9 \\
\midrule
\multirow{5}{*}{$\gamma_{11}$} 
  & 0 &  7.3 (0.3) &7.4 (0.31) & 7.4 (0.43) & 7.4 (0.38) & 7.3 (0.27)\\
  & 1 & \textbf{2.1 (0.89)} & \textbf{2.2 (0.97)} & \textbf{2.2 (1.2)} & \textbf{2 (0.87)} & \textbf{2 (0.82)} \\
  & 2 & \textbf{1.8 (0.72)} & \textbf{1.9 (0.76)} & \textbf{2 (0.71)} & \textbf{2 (0.78)} & \textbf{1.8 (0.72)} \\
  & $\mathrm{emp.}$ & 4.4 (1.3) & 4.4 (1.6) & 4.3 (1.6) & 4.4 (1.6) & 4.4 (1.6)\\
  & parametric ML & 1.2 (0.72) & 1.2 (0.87) & 1.3 (0.84) & 1.2 (0.8) & 1.1 (0.79)\\
  \midrule
\multirow{5}{*}{$\gamma_{12}$} 
  & 0 & 7.1 (0.36) &4.5 (0.81)  & \textbf{1.5 (1.8)} & \textbf{4.4 (0.84)} & 7.2 (0.43) \\
  & 1 & \textbf{2.1 (0.88)} & \textbf{1.6 (0.9)} & \textbf{0.64 (0.7)} & \textbf{1.5 (0.87)} &  \textbf{2.0 (0.89)} \\
  & 2 & \textbf{1.7 (0.68)} & \textbf{1.3 (0.58)} & \textbf{0.46 (0.5)} & \textbf{1.2 (0.61)} & \textbf{1.7 (0.70)} \\
  & $\mathrm{emp.}$ & 4.3 (1.4) & 4.3 (1.3) & 4.4 (1.6) & 4.4 (1.6) & 4.2 (1.3)\\
  & parametric ML & 0.46 (0.36) & 0.24 (0.15) & 0.11 (0.07) &  0.26 (0.17) & 0.45 (0.32) \\
  \bottomrule
\end{tabular}
\end{minipage}
\label{tab:mv_sim_results}
\end{table}
\renewcommand{\arraystretch}{1}

The results for $\gamma_{11}$ are reasonably constant across values of $\lambda_{12}$, and scale with $|\lambda_{12}|$ for $\gamma_{12}$ (with the exception of the empirical estimate). For all experiments, the results for $k \in \{1, 2\}$ are more accurate than the empirical estimator and bounded below by the parametric maximum likelihood estimator, as expected. Especially for $\gamma_{11}$, the $k \in \{1, 2\}$ models are comparable with the parametric maximum likelihood model, even only with a small number of fixed knots. The current study uses fixed knots; incorporating data-driven knot placement or adaptive basis selection may improve estimation further, particularly for capturing sharp or localised features in more complex models. These results confirm that our method provides a robust, flexible framework for estimating both marginal and cross-covariance functions, outperforming empirical estimates and approaching the performance of known fully parametric estimators.

\section{Application to spatio-temporal eddy formation} \label{sec:applications}

\begin{figure}[]
  \centering
  \includegraphics[width=\textwidth]{"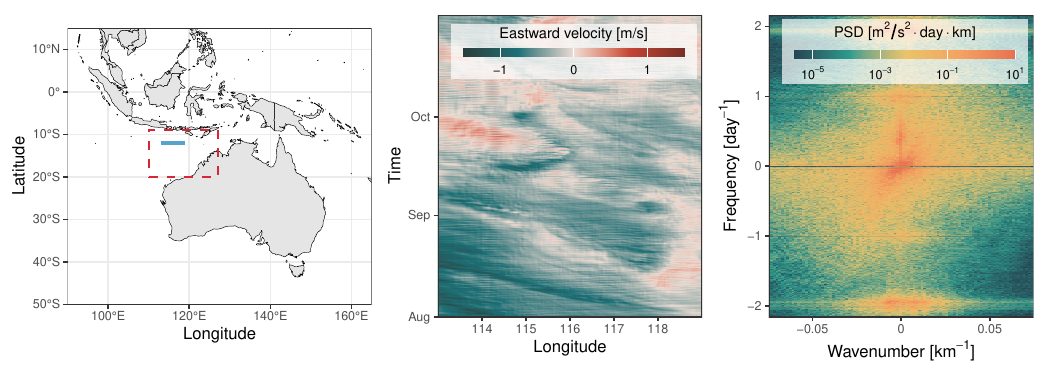"}
  \caption{SUNTANS model output over a 700\,km East--West transect. Left, the region of study; the model output is taken from the transect shown by the blue solid line. Middle, East--West velocity $u$ for a three-month period, showing low-frequency eddy structures. Right, periodogram of the complex velocity $w = u + \iota v$ over the full 12-month simulation; positive frequencies correspond to anticlockwise rotation and negative frequencies to clockwise rotation.}
  \label{fig:suntans_data}
\end{figure}

We conclude with an application that not only illustrates practical use of our method, but also validates its capacity to recover complex, physically interpretable structure in multidimensional spatio-temporal data. We study a complex spatio-temporal dataset from a high-resolution ocean simulation, generated using the Stanford Unstructured Nonhydrostatic Terrain-following Adaptive Navier-Stokes Simulator \citep[SUNTANS, ][]{rayson2025characteristic}. The simulation comprises $\sim$2.4 million observations spanning a 12-month period and resolves both tidal processes and large-scale eddies: slowly evolving, rotating structures generated by regional instabilities in the ocean circulation, that are a driver of extreme ocean currents. We focus on a $\sim$700\,km (6-degree) East--West transect of horizontal velocities, shown in the left panel of Figure~\ref{fig:suntans_data}; both the zonal (East--West) component $u$ and the meridional (North--South) component $v$ are available. This region was run with a high-resolution nested grid with the intention of modelling the ocean dynamics off Australia's North--West Shelf, a region of intense industry activity \citep[see][]{astfalck2018expert}. A three-month excerpt of $u$ is shown in the middle panel of Figure~\ref{fig:suntans_data}, and the rotary frequency--wavenumber periodogram of the complex velocity $w = u + \iota v$, computed over the full time series, is shown in the right panel \citep[see][for definition of the rotary spectrum]{gonella1972rotary}. The tidal processes are seen at the once (diurnal) and twice (semi-diurnal) per day marks on the frequency axis, and the low-frequency low-wavenumber power is associated with large-scale eddies. As is seen in the model-output, eddies tend to propagate westwards, suggesting some degree of non-separability in space and time.

We first remove a mean function of trend and tidal harmonics from the velocity components. For a real-valued process, the spectral density is even such that $f(\bomega) = f(-\bomega)$; for a complex-valued process this symmetry does not exist. The density of $w$ at positive frequencies corresponds to anticlockwise rotation. Thus any covariance model for $(u, v)$ with an even density cannot represent rotations. The method presented herein does not require any modifications, and simply a tensor-product basis is specified over the entire frequency--wavenumber plane. Our methodology provides a general framework for describing the ACF, and is agnostic to the choice of inference architecture. Owing to the construction in the frequency domain, Whittle-type inference methods are particularly natural \citep[see][]{dahlhaus1988small,sykulski2019debiased}. Further, given the size of this dataset the exact Gaussian likelihood \eqref{eqn:gaussian_lik} is unavailable to us, and we therefore maximise the Whittle log-likelihood \eqref{eqn:whittle}, both of which are defined and discussed in Section~\ref{sec:inference}.

\begin{figure}[]
  \centering
  \includegraphics[width=\textwidth]{"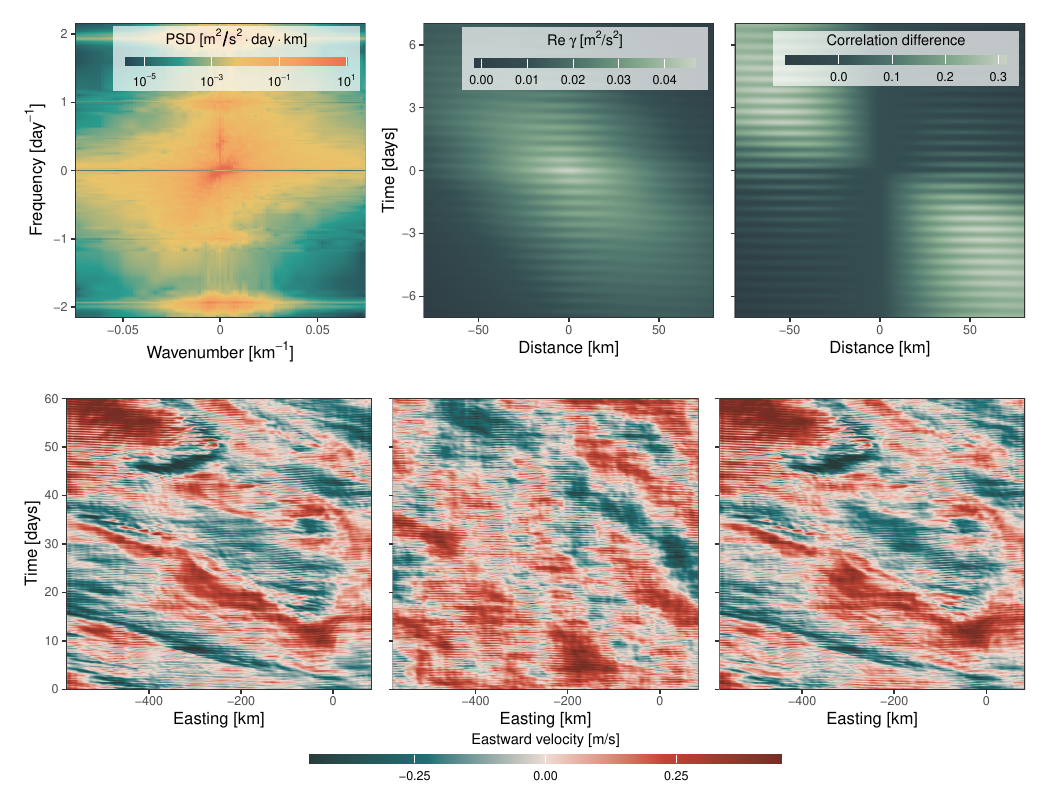"}
  \caption{Estimated PSD and ACF of the SUNTANS numerical model, and simulations from the fit. Top row: left, estimated rotary PSD of $w$ with degree $k=1$ tensor-product bases; middle, real part of the corresponding ACF computed in closed form; right, difference between the estimated non-separable ACF and a separable approximation based on outer products of the lag-zero axis profiles, normalised by variance.  Bottom row, the first 60 days of $u$, on a common colour scale: left, the model output; middle, an unconditional draw from the fit model; right, a draw from the fit spectrum with Fourier phases fixed to those of the model output.}
  \label{fig:suntans_fits}
\end{figure}

To model the spectral density, we apply our spline estimator using a tensor-product basis with $k = 1$. We place a geometric sequence of knots over frequency and wavenumber, and place additional knots surrounding known regions of anisotropy; in particular surrounding high-energy eddy signatures and tidal regions corresponding to the non-phase-locked tides; this gives 7683 basis functions over the whole plane. Inference proceeds by maximising the Whittle likelihood with respect to the basis coefficients, and took under a minute on a standard laptop. The left and middle panels of the top row of Figure~\ref{fig:suntans_fits} display the estimated PSD and the real part of the corresponding ACF, respectively. The estimated spectral density closely matches the periodogram, smoothing out the noise whilst preserving the tidal and low-frequency structure. The tidal component of the signal is represented in the ACF by the higher frequency perturbations, and the eddy signal is represented by the smoother mass rotated along the negative axis. This rotation reflects westwards eddy propagation. This propagation is oblique to the coordinate axes; the rotated basis of Appendix~\ref{sec:rotation} may better represent such structure. We retain the axis-aligned basis here and leave the rotated fit to future work. To explicitly assess the separability of the underlying spatio-temporal structure, we construct a separable approximation to the ACF by taking the outer product of the fit temporal and spatial autocovariance functions along their respective lag-zero axes. The right panel of Figure~\ref{fig:suntans_fits} shows the difference between the fit non-separable ACF and this separable surrogate, normalised in variance to highlight relative deviations. As expected, the two representations agree near the origin; however, substantial departures appear in the tail regions, where the separable model cannot replicate rotations in the ACF. This mismatch is typical of multidimensional processes where separability is incorrectly assumed \citep[e.g.][]{fuentes2008class}.

A fit spectral density fully specifies a stationary Gaussian field, so we may simulate from it directly. This requires an assumption of Gaussianity of the underlying field, which may be inappropriate, but serves here for an empirical evaluation. Assuming a periodic extension of the domain \citep[see, e.g.][]{guinness2017circulant} we may sample directly from the spectral density via an inverse fast Fourier transform of $\{\hat{f}_{\mathbf{k}}(\bomega) \, \delta\}^{1/2} Z(\bomega)$, with $Z(\bomega)$ independent standard complex Gaussian and $\delta$ the grid-cell area. The squared modulus of $Z(\bomega)$ is $\chi^2_2/2$ distributed corresponding to the standard fluctuations of the periodogram about $\hat{f}_{\mathbf{k}}(\bomega)$. The argument of $Z(\bomega)$ is uniform on $[0, 2\pi)$ and independent across frequencies. The bottom row of Figure~\ref{fig:suntans_fits} shows 60 days of $u$ from the model output (left) and from an unconditional draw (middle). The simulated draw has eddies of approximately the right size and propagation velocity, with the tidal oscillations superposed; however, the draw is a Gaussian approximation to dynamics known to be non-linear. Further, this is a random sample and so a visual comparison cannot say how much of the resemblance is due to $\hat{f}_{\mathbf{k}}$. To separate these contributions we re-simulate with fixed phases. In this sampling scheme, randomness enters via $Z$ and setting $Z = \hat{A}(\bomega) / \{\hat{f}_{\mathbf{k}}(\bomega) \, \delta\}^{1/2}$ with $\hat{A}$ the discrete Fourier transform of the demeaned model output, returns the demeaned model output exactly. This is analogous to drawing from a Gaussian process via $L z$, given a Cholesky matrix $L$, where the draw is determined by the seed that generates $z$. Keeping only the phase of $\hat{A}(\bomega)$, and drawing $\lvert Z(\bomega) \rvert^2 \sim \chi_2^2 / 2$ independently, results in a random draw with phases locked to the observed data. The result, in the right panel, has correlation 0.90 with the model output. Beyond serving as a validation device, fixing the phases to the data also reinstates certain structures that a stationary Gaussian assumption cannot replicate. The alignment of phases across frequencies and wavenumbers localises features in space and time; for instance, producing the non-linear sharp fronts at eddy boundaries. Fixing the phase field recovers these fronts, whereas the unconditional draw, with phases independent and uniform, cannot. 

This example highlights the flexibility of our method in capturing non-separable spatio-temporal dependencies. Crucially, no explicit mechanism is required to enforce non-separability; it emerges naturally from the data due to the structure of the model. The spline-based estimator yields interpretable representations with minimal prior specification or tuning, and when paired with appropriate inference architecture, remains computationally efficient even in multidimensional settings.

\section{Discussion} \label{sec:discussion}

Our results establish a flexible and theoretically grounded framework for non-parametric modelling of ACFs via spline-based expansions of the spectral density. We derive the analytic inverse Fourier transform of B-spline basis functions for any sequential knot design and to an arbitrary order of differentiability. Leveraging this result, we provide a constructive spline-based family that is uniformly dense among the ACFs of weakly stationary processes that possess a spectral density, and that accommodates univariate, multivariate, and multidimensional processes. This is supported by theory: we prove that our class of spline-based spectral densities is dense in the space of spectral densities of matrix-valued, weakly stationary ACFs with respect to the $L^1$-norm, so that the corresponding ACFs converge uniformly. Further, we provide a Jackson-type convergence inequality that upper-bounds the error of the modelled ACF in terms of the knot spacing and the spline degree. In the multivariate and multidimensional settings, in contrast to conventional separable models that impose strong structural constraints across components or dimensions, our spline kernel naturally accommodates non-separable dependencies. We demonstrate our method across two simulation studies, and an application to non-separable spatio-temporal ocean fields. These examples highlight the method's adaptability to both low- and high-dimensional data and show that interpretable ACFs can be recovered from data with minimal prior specification.

While our work provides a strong foundation, several important extensions remain. First, herein, inference proceeds via maximum likelihood estimates using the Gaussian and Whittle likelihoods. It would be interesting to study the methodology in a fully Bayesian framework. For instance, imposing our model as a prior in a GP regression. When the data become large, and the Gaussian likelihood (and posterior) becomes problematic to evaluate, it would be interesting to study how well our non-parametric model would fit under approximation strategies such as Vecchia's approximation. Second, we have largely discussed the knot placements as a modelling choice, but equally, knot number and locations may be estimated alongside the coefficients. This would fit naturally in a trans-dimensional Bayesian setting. Incorporating adaptive knot selection within the estimation routine would allow for data-driven refinement of spectral resolution and model complexity, especially in settings with localised features or heterogeneity. Third, the model developed here is stationary; however, the basis specification provides a natural route to verify this assumption. The eddy field in Section~\ref{sec:applications} may be considered stationary at the scale of the regional circulation, whereas over shorter time-frames a single eddy may be considered a transient feature over an otherwise quiescent field. The basis specification in our model separates the eddy scales from the quiescent periods. Refitting over a sequence of windows and comparing the band-limited coefficients can afford a diagnostic for stationarity that is indexed by frequency. Turning such a comparison into a formal test requires a null distribution for the coefficients under stationarity, which we do not develop here; methods such as segmented spectral models \citep{rosen2012adaptspec} and Fourier-domain tests of spatial stationarity \citep{bandyopadhyay2017test} provide a natural starting point.

In summary, we offer a unified, closed-form approach for modelling ACFs of stationary processes via spline kernels. The method provides both theoretical guarantees and practical flexibility, enabling principled, non-parametric inference in a broad range of applications. Future developments in adaptive refinement and alternative inferential strategies promise to further extend the reach of this framework.

\section*{Acknowledgements}

\if1\blind
{
  Dr. Lachlan Astfalck was in-part supported by the ARC ITRH for Transforming energy Infrastructure through Digital Engineering (TIDE), Grant No. IH200100009. Further thanks are to be given to Dr. Matt Rayson and Dr. William Edge for assistance with data curation.
} \fi

\if0\blind
{
  \textit{Redacted for double blind review.}
} \fi

\section*{Code and Data Availability}

\if1\blind
{
  Code is available at \texttt{github.com/astfalckl/bskernel}, including an R package implementation. A tutorial is provided with examples from both simulation studies. The dataset used in Section~\ref{sec:applications} is not public but may be made available upon request.
} \fi

\if0\blind
{
  Code is available at \textit{redacted for double blind review.} including an R package implementation. A tutorial is provided with examples from both simulation studies. The dataset used in Section~\ref{sec:applications} is not public but may be made available upon request.
} \fi

\subsection*{Declaration of use of Generative Artificial Intelligence}

This document was initially prepared using ChatGPT (models 4o and 5, via Codex) for language, grammar and spelling suggestions. In later revisions, the author used Claude (Anthropic, Opus 5.x) to verify mathematical results and simulation code. In particular, the revised proof of Theorem~3 was in part developed with this assistance. The research questions, methodology and conclusions are the author's own. The author has reviewed and edited all AI-assisted content and takes full responsibility for the content of the article.

\bibliographystyle{biometrika}
\bibliography{references}

\appendix

\section{Proofs and Derivations}

\subsection{Proofs to Theorems} \label{sec:proofs}

\begin{proof} (of Theorem~\ref{the:acf_closed_form})
  An outline of the proof to this Theorem is provided in Section~\ref{sec:non_uniform}. 
  Here, we complete the proof with the derivation of the analytical inverse Fourier transform of the truncated power basis in \eqref{eqn:double_truncation}, which underpins the general, non-uniform expression for \(\rho_{i,k}(\tau)\). For a fixed index \(j\), we wish to derive the inverse Fourier transform 
\[
\int_{-\infty}^\infty (\omega - \kappa_j)^k \cdot \one\{\omega \leq \kappa_{i+k+1}\} e^{2\pi \iota \omega \tau} \; \rd \omega = \int_{\kappa_j}^{\kappa_{i+k+1}} (\omega - \kappa_j)^k e^{2\pi \iota \omega \tau} \; \rd \omega.
\]
To evaluate this integral, let us perform a change of variable \( u = \omega - \kappa_j \), and define a constant $\lambda = 2\pi \iota \tau$, so that the integral becomes
\[
\int_0^{\kappa_{\mathrm{w},j}} u^k e^{\lambda (\kappa_j + u)} \; \rd u = e^{\lambda \kappa_j} \int_0^{\kappa_{\mathrm{w},j}} u^k e^{\lambda  u} \; \rd u,
\]
where we define \(\kappa_{\mathrm{w},j} = \kappa_{i+k+1} - \kappa_j\) as the width of the integration support.

Define the integral
\[
I_k(\lambda, \kappa_{\mathrm{w},j}) := \int_0^{\kappa_{\mathrm{w},j}} u^k e^{\lambda u} \, \rd u, \qquad \lambda \in \mathbb{C}, \quad \kappa_{\mathrm{w},j} > 0.
\]
For \(k = 0\), the integral is immediate,
\[
I_0(\lambda, \kappa_{\mathrm{w},j}) = \int_0^{\kappa_{\mathrm{w},j}} e^{\lambda u} \, \rd u = \frac{e^{\lambda \kappa_{\mathrm{w},j}} - 1}{\lambda}.
\]
We derive a closed-form expression for \(I_k\) using successive integration by parts which yields a recurrence
\begin{align*}
I_k(\lambda, \kappa_{\mathrm{w},j}) &= \left[ u^k \cdot \frac{e^{\lambda u}}{\lambda} \right]_0^{\kappa_{\mathrm{w},j}} - \frac{k}{\lambda} \int_0^{\kappa_{\mathrm{w},j}} u^{k-1} e^{\lambda u} \, \rd u \\
&= \frac{\kappa_{\mathrm{w},j}^k e^{\lambda \kappa_{\mathrm{w},j}}}{\lambda} - \frac{k}{\lambda} I_{k-1}(\lambda, \kappa_{\mathrm{w},j}).
\end{align*}
Iterating this \(k\) times yields,
\begin{align*}
I_k(\lambda, \kappa_{\mathrm{w},j}) &= \frac{e^{\lambda \kappa_{\mathrm{w},j}}}{\lambda} \left[ \kappa_{\mathrm{w},j}^k - \frac{k}{\lambda} \kappa_{\mathrm{w},j}^{k-1} + \frac{k(k-1)}{\lambda^2} \kappa_{\mathrm{w},j}^{k-2} - \cdots + (-1)^k \frac{k!}{\lambda^k} \right] - (-1)^k \frac{k!}{\lambda^{k+1}} \\
&= \frac{(-1)^k k!}{\lambda^{k+1}} \left( e^{\lambda \kappa_{\mathrm{w},j}} \sum_{l=0}^{k} \frac{(-\lambda \kappa_{\mathrm{w},j})^l}{l!} - 1 \right).
\end{align*}
Recalling the constant \( \lambda = 2\pi \iota \tau \), we obtain the desired inverse Fourier transform of each term in \eqref{eqn:double_truncation},
\[
\int_{\kappa_j}^{\kappa_{i+k+1}} (\omega - \kappa_j)^k e^{2\pi \iota \omega \tau} \, \rd \omega = \frac{(-1)^k k! e^{2\pi \iota \tau \kappa_j}}{(2\pi \iota \tau)^{k+1}} \left( e^{2\pi \iota \tau \kappa_{\mathrm{w},j}} \sum_{l=0}^{k} \frac{(-2\pi \iota \tau \kappa_{\mathrm{w},j})^l}{l!} - 1 \right).
\]

Since the B-spline basis \(B_{i,k}(\omega)\) is given as a linear combination of such truncated power functions via \eqref{eqn:double_truncation}, the corresponding ACF basis \(\rho_{i,k}(\tau)\) follows by linearity,
\begin{equation*}
  \rho_{i,k}(\tau) =\frac{(-1)^k \cdot k!}{(2 \pi \iota \tau)^{k+1}}  \sum_{j=i}^{i+k} \alpha_j e^{2 \pi \iota \kappa_j \tau} \left[
    e^{2\pi \iota \kappa_{\mathrm{w},j} \tau} \cdot \sum_{l=0}^k \frac{(-2\pi \iota \kappa_{\mathrm{w},j} \tau)^l}{l!} - 1
    \right],
\end{equation*}
recovering the full expression given in \eqref{eqn:non_uniform_acf_basis}.

\end{proof}

\begin{proof} (of Theorem~\ref{the:jackson})
  The proof to this Theorem proceeds in three parts: (1) a proof of Theorem XII(6) of \cite{de1978practical} of a Jackson-type inequality for uniform knots; (2) extension of this to non-uniform knots as in \citet{schumaker2007spline}; and (3) propagation of this error bound to define a convergence on the ACF. Theorem XII(6) of \cite{de1978practical}, states that for $f \in \mathcal{W}^{r,p}$ with $1 \leq r \leq k+1$ (de Boor indexes splines by their order, $k+1$) and a uniform knot placement on $[-a,a]$, such that $h = \kappa_i - \kappa_{i-1}$ is constant over $i$, there exists a spline $\hat{f} \in \mathcal{S}_{k,\mathcal{K}}$ such that
  \begin{equation*}
    \norm{f - \hat{f}}_{L^p([-a,a])} \leq C h^{r} \norm{f^{(r)}}_{L^p([-a,a])}.
  \end{equation*}
  This is proved by constructing a quasi-interpolant $\hat{f}_k = \Pi_h f = \sum_i \lambda_i(f) B_{i,k}$ where the $\lambda_i$ are local dual functionals, with $\lambda_i(B_{j,k}) = \delta_{ij}$ and $\lambda_i(f)$ depending only on $f$ over a knot interval in the support of $B_{i,k}$ \citep[see][Chapter~XII]{de1978practical}. As, $\Pi_h$ reproduces polynomials of degree $k$, such that $f - \Pi_h f = 0$ if $f \in \mathcal{P}_k$, for each $\omega \in [\kappa_{i}, \kappa_{i+1}]$, write
  \begin{equation*}
    f(\omega) = T_{r-1}(\omega) + R_r(\omega)
  \end{equation*}
  where $T_{r-1}(\omega)$ is the Taylor polynomial of degree $r-1 \leq k$, and $R_r(\omega)$ is the remainder term. As $\Pi_h T_{r-1} = T_{r-1}$, $f(\omega) - \Pi_h f(\omega) = (1 - \Pi_h)R_r(\omega)$ we may bound the norm over this interval as
  \begin{equation*}
    \norm{f(\omega) - \Pi_h f(\omega)}_{L^p([\kappa_i,\kappa_{i+1}])} \leq (1 + C_\Pi) \norm{R_r(\omega)}_{L^p([\kappa_i,\kappa_{i+1}])}
  \end{equation*}
  where $\norm{\Pi_h}_{L^p([\kappa_i,\kappa_{i+1}])} \leq C_\Pi$ from the operator norm. The remainder satisfies, 
  \begin{equation*}
    \norm{R_r(\omega)}_{L^p([\kappa_i,\kappa_{i+1}])} \leq C_R h^{r} \norm{f^{(r)}}_{L^p([\kappa_i,\kappa_{i+1}])}.
  \end{equation*} Thus, summing over the interval provides the global bound
  \begin{align*}
    \norm{f - \hat{f}}_{L^p([-a,a])}^p &= \sum_i \norm{f(\omega) - \Pi_h f(\omega)}_{L^p([\kappa_i,\kappa_{i+1}])}^p \\
    &\leq \sum_i C^p h^{rp} \norm{f^{(r)}}_{L^p([\kappa_i,\kappa_{i+1}])}^p \\
    &= C^p h^{rp} \norm{f^{(r)}}_{L^p([-a, a])}^p,
  \end{align*}
  and hence $\norm{f - \hat{f}}_{L^p([-a,a])} \leq C h^{r} \norm{f^{(r)}}_{L^p([-a, a])}$. For a non-uniform mesh size where $h_i = \kappa_{i+1} - \kappa_i$,
  \begin{align*}
    \norm{f - \hat{f}}_{L^p([-a,a])}^p &\leq \sum_i C^p h^{rp}_i \norm{f^{(r)}}_{L^p([\kappa_i,\kappa_{i+1}])}^p \\
    &\leq C^p h^{rp}_{\max} \norm{f^{(r)}}_{L^p([-a, a])}^p,
  \end{align*}
  and taking the $p$th root yields $\norm{f - \hat{f}}_{L^p([-a,a])} \leq C h^{r}_{\max} \norm{f^{(r)}}_{L^p([-a, a])}$. Here, the operator norm inequality $\norm{\Pi_h}_{L^p([\kappa_i,\kappa_{i+1}])} \leq C_\Pi$ is uniformly bounded only under the assumption that the mesh is quasi-uniform, that is, $h_{\max}/h_{\min} < \infty$ \citep[see][Chapter~6]{schumaker2007spline}. 
  
  Since $f$ is compactly supported to arbitrary precision, for every $\varepsilon_a > 0$ there exists some $a > 0$ such that
  \[
    \|f\|_{L^p(\mathbb{R} \setminus [-a,a])} < \varepsilon_a.
  \]
  The quasi-interpolant $\hat{f}_k = \Pi_h f$ is defined over the whole knot sequence, so that $\hat{f}_k \in \mathcal{S}_{k,\mathcal{K}}$ and $\hat{f}_k$ vanishes outside $[\kappa_0, \kappa_{m+k}]$. For each $B_{i,k}$ whose support extends beyond $[-a,a] = [\kappa_k, \kappa_m]$, we take the knot interval of $\lambda_i$ outside $[-a,a]$. Then $\hat{f}_k$ on $\mathbb{R} \setminus [-a,a]$ depends only on $f$ there and, enlarging $C_\Pi$ if necessary, $\|\hat{f}_k\|_{L^p(\mathbb{R} \setminus [-a,a])} \leq C_\Pi \|f\|_{L^p(\mathbb{R} \setminus [-a,a])}$. Consequentially,
  \begin{align*}
    \|f - \hat{f}_k\|_{L^p(\mathbb{R})} &\leq \|f - \hat{f}_k\|_{L^p([-a,a])} + \|f\|_{L^p(\mathbb{R} \setminus [-a,a])} + \|\hat{f}_k\|_{L^p(\mathbb{R} \setminus [-a,a])} \\
    &\leq C h_{\max}^{r} \|f^{(r)}\|_{L^p([-a,a])} + (1 + C_\Pi) \varepsilon_a.
  \end{align*}
  Note that $f \in L^p(\mathbb{R})$ by assumption, and that $\hat{f}_k$ is bounded with compact support and so lies in $L^p(\mathbb{R})$ for every $p$; the difference $f - \hat{f}_k$ therefore lies in $L^p(\mathbb{R})$ also. As the inverse Fourier transform in \eqref{eqn:bochner} is linear, $\gamma - \hat{\gamma}_k$ is the inverse Fourier transform of $f - \hat{f}_k$, and so the Hausdorff-Young inequality is applied below to the single function $f - \hat{f}_k$, rather than to $f$ and $\hat{f}_k$ separately.

  Finally, via the Hausdorff-Young inequality,
  \begin{equation*}
    \norm{\gamma - \hat{\gamma}}_{L^q(\mathbb{R})} \leq \norm{f - \hat{f}}_{L^p(\mathbb{R})} \leq C h^{r}_{\max} \norm{f^{(r)}}_{L^p([-a, a])} + (1 + C_\Pi) \varepsilon_a, \quad \text{for} \quad \frac{1}{p} + \frac{1}{q} = 1
  \end{equation*}
  and $1 \leq p \leq 2$. This is the stated bound with $C_{k,p} = \max(C, 1 + C_\Pi)$, which depends on $k$, $p$ and, through $C_\Pi$, on the bound on $h_{\max}/h_{\min}$.
\end{proof}

\begin{proof} (of Theorem~\ref{the:multivariate})
  Fix $k \geq 0$, and let $\bbf$ be the spectral density of $\bgamma \in \Lambda_M$, which is Hermitian and positive semi-definite for almost every $\omega$. For a knot sequence $\mathcal{K}$, define the linear averaging operator, which sets the coefficient of each $B_{i,k}$ to the local average of $g$ weighted by $B_{i,k}$,
  \[
    Q_{\mathcal{K}} g = \sum_{i=0}^{m-1} \frac{\int g B_{i,k} \, \rd\omega}{\int B_{i,k} \, \rd\omega} B_{i,k}, \qquad g \in L^1(\mathbb{R}; \mathbb{C}),
  \]
  such that $\hat{\bbf}_k = Q_{\mathcal{K}} \bbf$ entrywise with $\hat{\bbf}_k = \sum_i \textbf{C}_i B_{i,k}$ with $\textbf{C}_i = \int \bbf B_{i,k} \, \rd\omega / \int B_{i,k} \, \rd\omega$. As $B_{i,k}$ is real and $\bbf$ is Hermitian almost everywhere, $\textbf{C}_i^\mathrm{H} = \int \bbf^\mathrm{H} B_{i,k} \, \rd\omega / \int B_{i,k} \, \rd\omega = \textbf{C}_i$, so each $\textbf{C}_i$ is Hermitian. For any $\bv \in \mathbb{C}^M$, linearity of the integral gives
  \[
    \bv^\mathrm{H} \textbf{C}_i \bv = \sum_{r,s=1}^M \bar{v}_r v_s \frac{\int f_{rs} B_{i,k} \, \rd\omega}{\int B_{i,k} \, \rd\omega} = \frac{\int \bv^\mathrm{H} \bbf \bv \, B_{i,k} \, \rd\omega}{\int B_{i,k} \, \rd\omega}.
  \]
  As $\bv^\mathrm{H} \bbf(\omega) \bv \geq 0$ for almost every $\omega$ and $B_{i,k} \geq 0$, and the denominator $\int B_{i,k} \, \rd\omega = (\kappa_{i+k+1} - \kappa_i)/(k+1)$ is positive; thus $\bv^\mathrm{H} \textbf{C}_i \bv \geq 0$ for every $\bv$, and $\textbf{C}_i \succeq 0$. As $B_{i,k} \geq 0$, $\hat{\bbf}_k(\omega)$ is Hermitian and positive semi-definite for every $\omega$.

  For every knot sequence, $\hat{\bbf}_k = Q_{\mathcal{K}} \bbf$ is a member of our class, so density follows if $Q_{\mathcal{K}} g$ converges to $g$ in $L^1$ for every $g \in L^1(\mathbb{R}; \mathbb{C})$ as the knots are refined; applied to each entry of $\bbf$, this makes $\sum_{i,j} \norm{f_{ij} - \hat{f}_{ij,k}}_{L^1(\mathbb{R})}$ small, and the bound on the ACFs follows. In particular, we show that for every $g \in L^1(\mathbb{R}; \mathbb{C})$ and $\varepsilon > 0$ there exist $\delta > 0$ and a compact interval $S$ such that $\norm{g - Q_{\mathcal{K}} g}_{L^1} < 3\varepsilon$ for every knot sequence with $S \subseteq [\kappa_k, \kappa_m]$ and $h < \delta$, where $h = \max_j (\kappa_{j+1} - \kappa_j)$ is the maximum knot spacing. We use the following properties of B-splines \citep{de1978practical}: $B_{i,k} \geq 0$; $\sum_i B_{i,k} \leq 1$, with equality on $[\kappa_k, \kappa_m)$; and each $B_{i,k}$ is supported on an interval of length at most $(k+1)h$. 
  
  First, $Q_{\mathcal{K}}$ is bounded uniformly in $\mathcal{K}$, since
  \begin{equation} \label{eqn:th3_Q_bound}
    \norm{Q_{\mathcal{K}} g}_{L^1} \leq \sum_i \frac{|\int g B_{i,k} \, \rd\omega|}{\int B_{i,k} \, \rd\omega} \int B_{i,k} \, \rd\omega \leq \sum_i \int |g| B_{i,k} \, \rd\omega = \int |g| \sum_i B_{i,k} \, \rd\omega \leq \norm{g}_{L^1}.
  \end{equation}
  The first inequality is the triangle inequality, as $\norm{B_{i,k}}_{L^1} = \int B_{i,k} \, \rd\omega$ for the non-negative $B_{i,k}$; the second follows from $|\int g B_{i,k} \, \rd\omega| \leq \int |g| B_{i,k} \, \rd\omega$; the equality exchanges the finite sum and the integral; and the last inequality uses $\sum_i B_{i,k} \leq 1$.

  Second, let $g$ be continuous and supported in a compact interval $S \subseteq [\kappa_k, \kappa_m]$, with modulus of continuity $\pi_g(\delta) = \sup_{|\omega - \omega'| \leq \delta} |g(\omega) - g(\omega')|$. Then $g = g \sum_i B_{i,k}$ everywhere; on $[\kappa_k, \kappa_m)$ because $\sum_i B_{i,k} = 1$, outside $S$ because $g$ vanishes, and at $\kappa_m$ because $g(\kappa_m) = 0$ by continuity. Subtracting $Q_{\mathcal{K}} g(\omega)$ gives
  \begin{equation} \label{eqn:th3_average}
    g(\omega) - Q_{\mathcal{K}} g(\omega) = \sum_i B_{i,k}(\omega) \left\{ g(\omega) - \frac{\int g B_{i,k} \, \rd\omega'}{\int B_{i,k} \, \rd\omega'} \right\} = \sum_i B_{i,k}(\omega) \left\{ \frac{\int \{ g(\omega) - g(\omega') \} B_{i,k}(\omega') \, \rd\omega'}{\int B_{i,k} \, \rd\omega'} \right\}.
  \end{equation}
  Each $B_{i,k}$ is supported on an interval of length at most $(k+1)h$, hence $|\omega - \omega'| \leq (k+1)h$ throughout \eqref{eqn:th3_average}, and $|g(\omega) - g(\omega')| \leq \pi_g\{(k+1)h\}$. As the weights $B_{i,k}(\omega') / \int B_{i,k} \, \rd\omega'$ are non-negative and integrate to one, the bracketed term in \eqref{eqn:th3_average} is at most $\pi_g\{(k+1)h\}$. As $B_{i,k} \geq 0$ and $\sum_i B_{i,k} \leq 1$,
  \[
    |g(\omega) - Q_{\mathcal{K}} g(\omega)| \leq \sum_i B_{i,k}(\omega) \, \pi_g\{(k+1)h\} \leq \pi_g\{(k+1)h\}
  \]
  for every $\omega$. By bounding the region on which $g - Q_{\mathcal{K}} g$ can be non-zero, we can pass this result to the $L^1$ norm. As $g$ vanishes outside $S$, the coefficient $\int g B_{i,k} \, \rd\omega / \int B_{i,k} \, \rd\omega$ is non-zero only when the support of $B_{i,k}$ intersects $S$. The support of $B_{i,k}$ is at most of length $(k+1)h$ and lies within distance $(k+1)h$ of $S$ and hence $g - Q_{\mathcal{K}} g$ vanishes outside an interval of length $|S| + 2(k+1)h$. The $L^1$ norm is then at most the length of this interval times the pointwise bound on $|g - Q_{\mathcal{K}} g|$ within it, so 
  \begin{equation} \label{eqn:th3_step2_bound}
  \norm{g - Q_{\mathcal{K}} g}_{L^1} \leq \{|S| + 2(k+1)h\} \, \pi_g\{(k+1)h\}.
  \end{equation}
  As $g$ is continuous with compact support it is uniformly continuous, so $\pi_g(\delta) \to 0$ as $\delta \to 0$, and the bound tends to zero as $h \to 0$.
  
  Finally, the bound in \eqref{eqn:th3_step2_bound} requires $g$ to be continuous with compact support, which a spectral density need not be, e.g. if $g$ is discontinuous, $\pi_g(\delta)$ does not tend to zero. To extend the result for all $g \in L^1(\mathbb{R}; \mathbb{C})$ and $\varepsilon > 0$, we choose a continuous $g_\varepsilon$ supported in a compact interval $S$ such that $\norm{g - g_\varepsilon}_{L^1} < \varepsilon$. We apply the bound in \eqref{eqn:th3_step2_bound} to $g_\varepsilon$ and control the remaining difference with \eqref{eqn:th3_Q_bound}. Then,
  \[
    \norm{g - Q_{\mathcal{K}} g}_{L^1} \leq \underbrace{\norm{g - g_\varepsilon}_{L^1}}_{<\varepsilon, \text { by definition.}} + \underbrace{\norm{g_\varepsilon - Q_{\mathcal{K}} g_\varepsilon}_{L^1}}_{<\varepsilon, \text{ eqn. \eqref{eqn:th3_step2_bound}}} + \underbrace{\norm{Q_{\mathcal{K}}(g - g_\varepsilon)}_{L^1}}_{<\varepsilon, \text{ eqn. \eqref{eqn:th3_Q_bound}}} < 3 \varepsilon.
  \]
  Applying this to each of the $M^2$ entries of $\bbf$ with $\varepsilon = \varepsilon_\gamma / (3M^2)$, gives $\sum_{i,j} \norm{f_{ij} - \hat{f}_{ij,k}}_{L^1(\mathbb{R})} < \varepsilon_\gamma$. As $\gamma_{ij} - \hat{\gamma}_{ij,k}$ is the inverse Fourier transform of $f_{ij} - \hat{f}_{ij,k}$, and as $|e^{2\pi\iota\omega\tau}|=1$, via the triangle inequality $\sup_\tau |\gamma_{ij}(\tau) - \hat{\gamma}_{ij,k}(\tau)| \leq \norm{f_{ij} - \hat{f}_{ij,k}}_{L^1(\mathbb{R})}$, and $\hat{\Lambda}_{M,k}$ is uniformly dense in $\Lambda_M$.
\end{proof}

\subsection{Coefficients of the shifted truncated power function representation} \label{sec:coefficients}

As described in \eqref{eqn:tpb}, every B-spline \( B_{i,k}(\omega) \) of degree \(k\), with knot sequence \( \mathcal{K} = \{\kappa_0, \dots, \kappa_{m+k}\} \), may be written as a linear combination of shifted truncated power functions:
\[
B_{i,k}(\omega) = \sum_{j=i}^{i+k+1} \alpha_j (\omega - \kappa_j)^k_+.
\]
This representation is valid for arbitrary, non-uniform knot sequences. The coefficients \(\{\alpha_j\}\) depend only on the local configuration \(\{\kappa_i, \dots, \kappa_{i+k+1}\}\), and are nonzero only over this support. For any degree $k$, the coefficients are given by the divided difference \citep[see][]{de1978practical}
\[\alpha_j = (-1)^{k+1} (\kappa_{i+k+1} - \kappa_i) \Big/ \prod_{l = i, \, l \neq j}^{i+k+1} (\kappa_j - \kappa_l), \qquad j = i, \dots, i+k+1.\]
We also give the explicit expressions for \(k \leq 2\).

\paragraph{Linear B-splines.}

Given knots \(\kappa_i < \kappa_{i+1} < \kappa_{i+2}\), the B-spline basis can be written equivalently in the truncated power basis form and piecewise polynomial form, respectively, as
\begin{align*}
B_{i,1}(\omega) &= \alpha_i (\omega - \kappa_i)^1_+ + \alpha_{i+1} (\omega - \kappa_{i+1})^1_+ + \alpha_{i+2} (\omega - \kappa_{i+2})^1_+ \\
&= \begin{cases}
  \displaystyle \frac{\omega - \kappa_i}{\kappa_{i+1} - \kappa_i}, & \omega \in [\kappa_i, \kappa_{i+1}), \\[1.5ex]
  \displaystyle \frac{\kappa_{i+2} - \omega}{\kappa_{i+2} - \kappa_{i+1}}, & \omega \in [\kappa_{i+1}, \kappa_{i+2}), \\[1.5ex]
  0, & \text{otherwise}.
  \end{cases}
\end{align*}
Note, the third term $\alpha_{i+2} (\omega - \kappa_{i+2})^1_+$ ensures $B_{i,1}(\omega) = 0$ for $\omega \geq \kappa_{i+2}$. Solving for the coefficients gives
\begin{align*}
  \alpha_i = \frac{1}{\kappa_{i+1} - \kappa_i}, \quad
  \alpha_{i+1} = -\left( \frac{\kappa_{i+1} - \kappa_i}{\kappa_{i+2} - \kappa_{i+1}} + 1 \right) \cdot \alpha_i, \quad
  \alpha_{i+2} = -\alpha_i - \alpha_{i+1}.
  \end{align*}

\paragraph{Quadratic B-splines.}

Given knots \(\kappa_i < \kappa_{i+1} < \kappa_{i+2} < \kappa_{i+3}\), the quadratic B-spline basis function \(B_{i,2}(\omega)\) can be written equivalently in truncated power basis form and piecewise polynomial form, respectively, as
\begin{align*}
B_{i,2}(\omega) &= \alpha_i (\omega - \kappa_i)^2_+ + \alpha_{i+1} (\omega - \kappa_{i+1})^2_+ + \alpha_{i+2} (\omega - \kappa_{i+2})^2_+ + \alpha_{i+3} (\omega - \kappa_{i+3})^2_+ \\
&= \begin{cases}
  \displaystyle \frac{(\omega - \kappa_i)^2}{(\kappa_{i+2} - \kappa_i)(\kappa_{i+1} - \kappa_i)}, & \omega \in [\kappa_i, \kappa_{i+1}), \\[1.5ex]
  \displaystyle \frac{(\omega - \kappa_i)(\kappa_{i+2} - \omega)}{(\kappa_{i+2} - \kappa_i)(\kappa_{i+2} - \kappa_{i+1})} +
  \frac{(\kappa_{i+3} - \omega)(\omega - \kappa_{i+1})}{(\kappa_{i+3} - \kappa_{i+1})(\kappa_{i+2} - \kappa_{i+1})}, &  \omega \in [\kappa_{i+1}, \kappa_{i+2})\\[1.5ex]
  \displaystyle \frac{(\kappa_{i+3} - \omega)^2}{(\kappa_{i+3} - \kappa_{i+2})(\kappa_{i+3} - \kappa_{i+1})}, & \omega \in [\kappa_{i+2}, \kappa_{i+3}), \\[1.5ex]
  0, & \text{otherwise}.
\end{cases}
\end{align*}
As in the linear case, the inclusion of the final term \(\alpha_{i+3}(\omega - \kappa_{i+3})^2_+\) ensures compact support on \([\kappa_i, \kappa_{i+3}]\). Solving for the coefficients gives
\begin{align*}
\alpha_i &= \frac{1}{(\kappa_{i+1} - \kappa_i)(\kappa_{i+2} - \kappa_i)}, \\
\alpha_{i+1} &= -\frac{\kappa_{i+3} - \kappa_i}{(\kappa_{i+1} - \kappa_i)(\kappa_{i+2} - \kappa_{i+1})(\kappa_{i+3} - \kappa_{i+1})}, \\
\alpha_{i+2} &= \frac{ -\alpha_i (\kappa_{i+3} - \kappa_i)^2 - \alpha_{i+1} (\kappa_{i+3} - \kappa_{i+1})^2 }{(\kappa_{i+3} - \kappa_{i+2})^2}, \\
\alpha_{i+3} &= -(\alpha_{i} + \alpha_{i+1} + \alpha_{i+2}).
\end{align*}

\section{Extensions of the tensor-product basis} \label{app:adaptivity}

The tensor-product basis in Section~\ref{sec:multidimensional} is limited in two ways. First, it is not locally adaptive. When the knots are fixed this is of little consequence; however, when knots are refined adaptively or estimated, inserting a single knot in the $j$th direction introduces $\prod_{l \neq j} m_l$ new basis functions, which limits the construction in high dimensions and in spatially heterogeneous settings. Appendices~\ref{sec:hierarchical} and~\ref{sec:t_mesh} set out two constructions from the computer graphics literature, hierarchical B-splines and T-mesh splines, that refine the basis locally; Figure~\ref{fig:mesh} compares their meshes with the tensor-product grid. Second, every basis function is supported on a box aligned with the coordinate axes (including the extensions in \ref{sec:hierarchical} and~\ref{sec:t_mesh}). Appendix~\ref{sec:rotation} extends the methodology to allow for rotated bases.

\begin{figure}[pos=H]
\centering
\begin{tikzpicture}[scale=1.2]

\begin{scope}
  \foreach \x in {0,1,2,3} {
    \draw[gray] (\x,0) -- (\x,3);
    \foreach \y in {0,1,2,3} {
      \draw[gray] (0,\y) -- (3,\y);
      \filldraw[black] (\x,\y) circle (1.25pt);
    }
  }
  \node at (1.5,-0.5) {\small Tensor-product grid};
\end{scope}

\begin{scope}[shift={(8,0)}]
  \foreach \y in {0,2,3} {
    \draw[gray] (0,\y) -- (3,\y);
    \foreach \x in {0,1,3} {
      \filldraw[black] (\x,\y) circle (1.25pt);
    }
  }
  \draw[gray] (0,1) -- (1,1);
  
  \draw[gray] (0,0) -- (0,3);
  \draw[gray] (1,0) -- (1,3);
  \draw[gray] (2,2) -- (2,3); %
  \draw[gray] (3,0) -- (3,3);

  \filldraw[black] (2,3) circle (1.25pt);
  \filldraw[black] (0,1) circle (1.25pt);

  \draw[blue, dashed, line width=0.4mm] (0,1) -- (3,1); %
  \draw[blue, dashed, line width=0.4mm] (1,0) -- (1,2); %
  \node[blue] at (0,1) {\Large $\ast$};
  \node[blue] at (1,0) {\Large $\ast$};
  \node[blue] at (3,1) {\Large $\ast$};
  \node[blue] at (1,2) {\Large $\ast$};

  \filldraw[red] (2,2) circle (2pt); %
  \filldraw[red] (1,1) circle (2pt); %
  \node at (1.5,-0.5) {\small T-mesh};
\end{scope}

\begin{scope}[shift={(4,0)}]
  \foreach \x in {0,1,2,3} {
    \draw[gray] (\x,0) -- (\x,3);
    \draw[gray] (0,\x) -- (3,\x);
  }
  \foreach \x in {1.5,2,2.5,3} {
    \draw[gray] (\x,1.5) -- (\x,3);
  }
  \foreach \x in {1.5,3} {
    \draw[gray] (\x,1.5) -- (\x,3);
  }
  \foreach \y in {1.5,2,2.5,3} {
    \draw[gray] (1.5,\y) -- (3,\y);
  }
  \foreach \y in {1.5,,3} {
    \draw[gray] (1.5,\y) -- (3,\y);
  }
  \foreach \x in {1.5,1.75,2,2.25} {
    \draw[gray] (\x,1.5) -- (\x,2.25);
  }
  \foreach \y in {1.5,1.75,2, 2.25} {
    \draw[gray] (1.5,\y) -- (2.25,\y);
  }

  \draw[line width=0.4mm, black] (0,0) -- (0,3);
  \draw[line width=0.4mm, black] (0,0) -- (3,0);
  \draw[line width=0.4mm, black] (0,3) -- (3,3);
  \draw[line width=0.4mm, black] (3,0) -- (3,3);
  \draw[line width=0.4mm, black] (1.5,1.5) -- (1.5,3);
  \draw[line width=0.4mm, black] (1.5,1.5) -- (3,1.5);
  \draw[line width=0.4mm, black] (1.5,3) -- (3,3);
  \draw[line width=0.4mm, black] (3,1.5) -- (3,3);
  \draw[line width=0.4mm, black] (1.5,1.5) -- (1.5,2.25);
  \draw[line width=0.4mm, black] (1.5,1.5) -- (2.25,1.5);
  \draw[line width=0.4mm, black] (1.5,2.25) -- (2.25,2.25);
  \draw[line width=0.4mm, black] (2.25,1.5) -- (2.25,2.25);

  \foreach \x in {0,1,2,3} {
  \foreach \y in {0,1,2,3} {
    \fill[black] (\x,\y) circle (1.25pt);
  }
}

  \foreach \x in {1.5,2,2.5,3} {
    \foreach \y in {1.5,2,2.5,3} {
      \fill[black] (\x,\y) circle (1.25pt);
    }
  }
  \foreach \x in {1.5,1.75,2, 2.25} {
    \foreach \y in {1.5,1.75,2, 2.25} {
      \fill[black] (\x,\y) circle (1.25pt);
    }
  }
  \node at (1.5,-0.5) {\small Hierarchical Grid};
\end{scope}

\end{tikzpicture}
\caption{Comparison of a regular tensor-product grid (left), a hierarchical grid with three hierarchy levels (middle), and a T-mesh with two T-junctions and an example knot cross (right). Dots denote mesh vertices or knot locations.}
\label{fig:mesh}
\end{figure}

\subsection{Hierarchical splines} \label{sec:hierarchical}

Hierarchical B-splines offer a principled mechanism for local refinement that preserves the compact support, smoothness, and analytic tractability of classical B-splines while avoiding the exponential growth
associated with tensor-product grids. We provide a brief overview, see \cite{vuong2011hierarchical} or \cite{kunoth2018foundations} for more rigorous development. The key idea is to construct a nested sequence of spline spaces \(\mathcal{S}^0 \subset \mathcal{S}^1 \subset \cdots \subset \mathcal{S}^L\), where each level \(\ell\) corresponds to a spline space of degree \(k\) defined over a dyadic refinement of the knot grid. Let \(\mathcal{K}_\ell\) denote the knot set at level \(\ell\), obtained by uniformly subdividing the previous grid, that is, \(\mathcal{K}_{\ell+1}\) is a refinement of \(\mathcal{K}_\ell\) such that \(\max_i (\kappa_{i+1}^{(\ell+1)} - \kappa_i^{(\ell+1)}) = \frac{1}{2} \max_i (\kappa_{i+1}^{(\ell)} - \kappa_i^{(\ell)})\). The corresponding spline space \(\mathcal{S}^\ell = \mathcal{S}_{k, \mathcal{K}_\ell}\) has a B-spline basis \(\{B^{(\ell)}_{\textbf{i},k}(\bomega)\}\), and given nested regions $\Omega_0 \supseteq \Omega_1 \supseteq \cdots \supseteq \Omega_L$ of the frequency domain, marking where each level of resolution is wanted, with $\Omega_{L+1} = \emptyset$, we define a hierarchical basis by selecting those B-splines at level \(\ell\) whose support lies in $\Omega_\ell$ but not in $\Omega_{\ell+1}$ \citep{vuong2011hierarchical}. 

From a spectral modelling perspective, the hierarchical B-spline basis allows resolution to vary adaptively in spectral space, enabling fine detail where needed while coarsely representing smooth or uninformative regions. As each basis function \(\{B^{(\ell)}_{\textbf{i},k}(\bomega)\}\) is still a scaled B-spline with compact support, its inverse Fourier transform can be computed analytically, as previously given. The resulting ACF approximation takes the form
\[
\hat{\gamma}_{\mathrm{HB}}(\btau) = \sum_{\ell = 0}^L \sum_{\textbf{i} \in \mathcal{A}_\ell} c_\textbf{i}^{(\ell)} \rho^{(\ell)}_{\textbf{i},k}(\btau),
\]
where \(\rho^{(\ell)}_{i,k}(\btau)\) is the ACF basis associated with \(B^{(\ell)}_{\textbf{i},k}(\bomega)\), and \(\mathcal{A}_\ell\) is the set of active indices at level \(\ell\). 
A visualisation of three levels of hierarchical refinement is shown in the middle panel of Figure~\ref{fig:mesh}.

\subsection{T-mesh splines} \label{sec:t_mesh}

T-mesh splines provide a more flexible alternative to hierarchical splines, allowing for local refinement without the combinatorial overhead of tensor-product constructions. A T-mesh generalises a regular grid by permitting T-junctions, locations where a row or column terminates without fully bisecting the domain; see the right panel of Figure~\ref{fig:mesh} for an example mesh with two T-junctions. This structure enables the insertion of new knots locally, without inducing cross-dimensional refinement, as illustrated in the right panel of Figure~\ref{fig:mesh}. In the T-mesh literature, control points are often referred to as anchors; we retain the term \emph{knot} for consistency with our earlier notation.

For notational simplicity we illustrate in \( D = 2 \) dimensions; higher $D$ follows analogously. Let \(\bm{\omega}_{i,j} = (\omega_{1,i}, \omega_{2,j}) \in \mathbb{R}^2\) denote a local knot. Each such knot defines a tensor-product B-spline basis function constructed from a local \emph{knot cross}, a collection of nearby univariate knot positions along each axis. We denote these neighbourhoods as
\[
\mathcal{K}^{(1)}_{i,j} = \{\omega_{1,i - \lceil k/2 \rceil}, \dots, \omega_{1,i + \lfloor k/2 \rfloor + 1}\}, \quad
\mathcal{K}^{(2)}_{i,j} = \{\omega_{2,j - \lceil k/2 \rceil}, \dots, \omega_{2,j + \lfloor k/2 \rfloor + 1}\},
\]
where the superscript indicates the coordinate direction and the subscript \((i,j)\) highlights that these knot vectors are local to the knot \(\bm{\omega}_{i,j}\). In effect, the knot cross is a directional stencil of \(k+2\) knots surrounding \(\bm{\omega}_{i,j}\); an example is shown in blue for the lower-left T-junction in Figure~\ref{fig:mesh}. Not all T-meshes yield linearly independent bases, to guarantee stability, the mesh must be analysis-suitable \citep[see][]{li2014analysis}.

The corresponding basis function is defined as
\[
B_{[i,j],k}(\bm{\omega}) = B_{i,k}^{(1)}(\omega_1; \mathcal{K}^{(1)}_{i,j}) \cdot B_{j,k}^{(2)}(\omega_2; \mathcal{K}^{(2)}_{i,j}),
\]
where each \(B_{i,k}^{(r)}(\cdot; \mathcal{K})\) is a univariate B-spline of degree \(k\), defined over the extracted knot vector \(\mathcal{K}\) in direction \(r\). These basis functions retain the essential properties of classical B-splines, and  anisotropy is naturally encoded through the directional spacing in the \(\mathcal{K}^{(r)}_{i,j}\). As in the tensor-product case, the inverse Fourier transform of each basis function remains separable, and the corresponding autocovariance components are given by products of univariate inverse transforms as in Theorem~\ref{the:acf_closed_form}.

\subsection{Rotated bases} \label{sec:rotation}

The supports of the tensor-product basis, and of the two extensions in Appendices~\ref{sec:hierarchical} and~\ref{sec:t_mesh}, have faces normal to the coordinate axes. This may cause inefficient estimation (in the number of bases required) when representing features oblique to the axes such as the low-frequency low-wavenumber eddy fields in Section~\ref{sec:applications}. Instead, by evaluating each basis at a linearly transformed frequency, we may form a family of rotated bases that remains closed-form. Let $A \in \mathbb{R}^{D \times D}$ be invertible and let $\rho_{\mathbf{i}, \mathbf{k}}$ be the ACF basis function of \eqref{eqn:multivariate_rho}. Then
\[
    \int_{\mathbb{R}^D} B_{\mathbf{i}, \mathbf{k}}(A \bomega) \, e^{2 \pi \iota \bomega^\T \btau} \, \rd \bomega = \lvert \det A \rvert^{-1} \, \rho_{\mathbf{i}, \mathbf{k}}(A^{-\T} \btau),
  \]
which is $\rho_{\mathbf{i}, \mathbf{k}}(A \btau)$ when $A = R_{\bm{\theta}}$ is a rotation as $\lvert \det A \rvert = 1$ and $A^{-\T} = A$. The fit ACF is then the axis-aligned ACF \eqref{eqn:multidimensional} evaluated at $R_{\bm{\theta}} \btau$. For fixed $\bm{\theta}$ the estimation problem in $c$ is unchanged, so $\bm{\theta}$ may be treated as a constant in the likelihood.

We may also let the angles $\bm{\theta}_{\mathbf{i}}$ vary with $\mathbf{i}$. By linearity, a separate invertible $A_{\mathbf{i}}$ may be assigned to each basis function, giving
\begin{equation} \label{eqn:rotated_psd}
  \hat{f}_{\mathbf{k}}(\bomega) = \sum_{\mathbf{i}} c_{\mathbf{i}} B_{\mathbf{i}, \mathbf{k}}(A_{\mathbf{i}} \bomega),
  \qquad
  \hat{\gamma}_{\mathbf{k}}(\btau) = \sum_{\mathbf{i}} c_{\mathbf{i}} \lvert \det A_{\mathbf{i}} \rvert^{-1} \rho_{\mathbf{i}, \mathbf{k}}(A_{\mathbf{i}}^{-\T} \btau).
\end{equation}
The ACF remains closed form, by \eqref{eqn:rotated_psd}, and it can combine components of different orientation, which in general no rotation of an axis-aligned ACF can represent. It contains the fixed rotation $R_{\bm{\theta}}$ as a special case, and so approximates at least as well. However, the per basis function angles enter the likelihood non-linearly, and in practice with finite samples would require some degree of regularisation across neighbouring bases. Here, physical motivation can be used to inform each of the angles. For example, we may consider a shear in space-time whereby we write $\bomega = (\bomega_s, \omega_t)$ for spatial frequency $\bomega_s \in \mathbb{R}^{D-1}$ and temporal frequency $\omega_t$, and let $A(\bomega_s, \omega_t) = (\bomega_s, \omega_t + \bv^\T \bomega_s)$ for a velocity $\bv \in \mathbb{R}^{D-1}$. This represents the ACF of an axis-aligned field advected with velocity $\bv$.

\end{document}